\documentclass[11pt,a4paper]{article}

\usepackage[left=20mm,top=20mm,right=20mm,bottom=18mm]{geometry}

\usepackage{amsmath, amsthm, amsfonts}
\usepackage{natbib}
\setcitestyle{authoryear,open={(},close={)}}
\newcommand{\textcite}[1]{\cite{#1}}
\usepackage[colorlinks=true,urlcolor=blue,citecolor=blue]{hyperref}
\usepackage{overpic}
\usepackage{relsize}
\usepackage{booktabs}

\allowdisplaybreaks
\usepackage{xcolor}

\usepackage[scaled]{helvet}
\makeatletter
\renewcommand{\paragraph}{%
  \@startsection{paragraph}%
  {4}
  {\z@}
  {0.5ex \@plus.2ex \@minus.2ex}
  {-1em}
  {\normalfont\normalsize\itshape}
}
\makeatother

\usepackage{xr} 
\usepackage{comment}
\excludecomment{journal}
\includecomment{plain}

\newcommand{\rev}[1]{#1}
\newcommand{\revii}[1]{#1}
\newcommand{\reviii}[1]{#1}

\begin{document}

\title{
A large population of cell-specific action potential models replicating fluorescence recordings of voltage in rabbit ventricular myocytes} 
\author{Radostin D Simitev\,$^{1,\ast}$,  Rebecca J Gilchrist\,$^{2}$,
  Zhechao Yang$^1$ \\[4pt] Rachel  Myles\,$^{2}$, Francis Burton\,$^{2}$ and Godfrey L Smith\,$^2$ \\[5mm]
  {\small
    $^1$ School of Mathematics \& Statistics, University of
    Glasgow, Glasgow, UK} \\
  {\small  $^2$ School of Cardiovascular \& Metabolic Health
    University of Glasgow, Glasgow, UK }\\
  {\small  $^\ast$ Corresponding author:
    \href{mailto:Radostin.Simitev@glasgow.ac.uk}{Radostin.Simitev@glasgow.ac.uk},
    \href{https://orcid.org/0000-0002-2207-5789}{orcid.org/0000-0002-2207-5789}}}
\date{Accepted for publication in Royal Society Open Science
  (ISSN:2054-5703) on 2025-01-13.\\
Supplementary material available as open source from Royal Society Open Science.}
\maketitle

\graphicspath{.}

\newcommand{\correct}[1]{#1}
\begin{plain}
\newcommand{\keywords}[1]{\textbf{Keywords:} #1}
\end{plain}

\begin{abstract}
Recent high-throughput experiments unveil substantial
electrophysiological diversity among uncoupled healthy myocytes under
identical conditions.
To quantify inter-cell variability, the values of a subset of the
parameters  in a well-regarded mathematical model of the action
potential of rabbit ventricular myocytes are estimated from
fluorescence voltage measurements of a large number of cells. Statistical
inference yields a population of nearly 1200 cell-specific model
variants \rev{that, on a population-level replicate experimentally
measured biomarker ranges and distributions, and in contrast to
earlier studies, also match experimental biomarker values on
a cell-by-cell basis. This model population} may be regarded as a random  sample from
the phenotype of healthy rabbit ventricular myocytes. Uni-variate and
bi-variate joint marginal distributions of the 
estimated parameters are presented, and the parameters dependencies of
several commonly utilised electrophysiological biomarkers are determined.
Parameter values are weakly correlated, while summary metrics such
as the action potential duration are not strongly dependent on any
single electrophysiological characteristic of the myocyte. 
Our results demonstrate the feasibility of accurately and efficiently
fitting entire action potential waveforms at scale.

\end{abstract}

\begin{journal}
\begin{fmtext}
\end{journal}
\section{Introduction}
\label{sec01}
\paragraph{Cellular variability.}
Genetically identical cardiomyocytes, even ones that have developed 
in identical extracellular conditions, exhibit differences in
their electrophysiological properties.
Inter-cell variability has been confirmed using a range of
experimental techniques in cardiac
tissues from various species 
\citep{Antzelevitch1991,Feng1998,Britton2013,Sanchez2014,Passini2016,Zhou2016}. 
Inter-cell variability affects physiological function upstream at tissue and organ
levels with one clinically important effect being that 
a drug therapy designed to inhibit specific ion channel(s) will have
different outcomes on different members of the
population \citep{Niepel2009,Yang2012}.
Experimental determination of the differences in multiple ionic
conductances underlying this inter-cell heterogeneity is not feasible
and so it is necessary to examine them in mechanistic mathematical
models. 

\paragraph{Mainstream approaches to modelling of variability.}
Advances in the mathematical and statistical
modelling of cardiomyocyte variability 
are summarised in \citep{Ni2018,Lei2020,Whittaker2020}, with 
two main strategies emerging:
a ``population-based'' approach
and a ``sample-specific'' approach \citep{Ni2018}. Both begin with an
appropriate generic cardiomyocyte action potential model as a baseline.
Generic models for various species and cardiac tissues have been
extensively developed over the past 70 years \citep{Winslow2011,Amuzescu2021}. 
Population-based approaches proceed to generate model 
variants of their baseline model by randomly sampling parameter values from
hypothetical distributions.
In some studies, further
rejection-subsampling is performed to fit
\begin{journal}
\end{fmtext}
\maketitle
\newpage
\noindent
\end{journal}
presumed or experimental
distributions of one or more action  potential characteristics
(biomarkers). Examples include the works
\citep{Romero2009,Britton2013,Muszkiewicz2016,Morotti2017}. 
Sample-specific modelling
approaches re-estimate the parameter 
values of their baseline
model using cell-specific datasets. Examples
include the studies
\citep{Dokos2004,Syed2005,Bot2012,Groenendaal2015,KroghMadsen2016,Aziz2022}. 
Population-based approaches offer the advantage of statistically
meaningful population sizes of the order of $10^4$
randomised models. Such sizes are larger than the number of cells  easily
measurable in experiments, yield more accurate
descriptive statistics and facilitate detection of effects that may
go unnoticed in small-size populations.
The critical drawback of population-based approaches 
lies in the lack of direct correspondence between individual variants
from the model population and individual biological cells from the pool
of cardiomyocytes measured. This leads to model populations that do
not align with biomarker distributions which they have not been calibrated
to fit beforehand and undermines their predictive capacity.
In contrast, sample-specific approaches are attractive for their
one-to-one correspondence between biological cells and tailored mathematical models.
Customised models facilitate quantification of the internal state of
specific cells, calculation of characteristics that are either
unmeasured or difficult to assess, and prediction of behavior under
diverse external conditions.    
However, developing sample-specific models is resource-intensive,
requiring ample experimental data to constrain the models and
substantial effort for subsequent parameter estimation. Consequently,
this approach has been primarily limited to single or very few cells
and rarely employed to study inter-cell variability. 

\paragraph{Goals.}
The general goal of this article is to combine the advantages of
sample-specific and  population-based
approaches. This has now become possible due to increases in
computing power and gradual improvement in classical and probabilistic
algorithms for parameter estimation \citep{Arridge2019}. More
importantly, this has been enabled by recent advances in optics-based
techniques for cardiac electrophysiology  \citep{Mllenbroich2021} 
that make it possible to  develop high-throughput automatic
and semi-robotic platforms capable of recording transmembrane voltage
traces in several thousand uncoupled cardiomyocytes per hour 
\citep{Warren2010,Herron2012,Heinson2023,Lee2023}.
The particular aim of the study is to develop and apply such a hybrid
cell-specific population-based methodology for statistical analysis
and interpretation of new experimental recordings of cardiomyocyte
action potential waveforms from our group.
In our recent work \citep{Lachaud2022}, voltage-sensitive
fluorescent dyes were employed to capture action potential waveforms
from nearly 500 cardiomyocytes isolated from the left ventricular wall
of 12 rabbit hearts, revealing substantial variability among cells.  
In particular, the durations of action potentials at 90\%
repolarisation (APD$_{90}$) had wide (40-50 ms) inter-quartile
ranges indicating variability
considerably exceeding that of median APD$_{90}$ values across
different animal hearts and within the endo-epicardial and
apical-basal regions of each heart. 
A conventional population-based analysis of the experiment
of \citep{Lachaud2022} was then performed. The detailed ionic
current model of \textcite{Shannon2004} was selected as a baseline mathematical
representation of the rabbit myocytes, 50,000 model variants with randomly
sampled values of eight sensitive parameters were then generated, and the
population was calibrated by rejecting variants falling outside the
experimentally measured histogram distribution of 
APD$_{90}$ values.
However, a large amount of valuable experimental information was
discarded in the process. Notably, while complete AP traces 
comprising voltage values recorded at a frequency of 10 kHz
(i.e.,10,000 voltage values per second) were available for all cells,
only a single value (APD$_{90}$) per cell was utilised.
Model variants were not constrained to reproduce cell-specific
APD$_{90}$ values, only to return values consistent with an
experimental-like histogram distribution of this
biomarker. Consequently, the model population did not reproduce the
experimentally measured distributions of other measured biomarkers,
e.g.{} APD$_{30}$ and APD$_{50}$.
In the present article, we attempt to improve on this approach
by constraining parameter values to find cell-specific model variants
that reproduce the entire action potential waveforms of
individual biological myocytes. We will use a large population of over
1200 myocytes measured for this purpose following optics based
experimental protocols similar to that of \citep{Lachaud2022}. This 
refined approach ensures a more comprehensive utilisation of the
experimental data, enhancing the 
accuracy and robustness of our analyses and the predictive capacity of
both individual models and the overall model population.

\section{Methodology}
\label{sec02}
To construct an ensemble of cell-specific action potential models
we estimate an individualised set of parameter values
for each experimentally measured cardiomyocyte.
Various inference approaches exist, frequentist as well as Bayesian.
Here we outline our chosen methodology with associated assumptions  and  notation.

\subsection{Parameter estimation in dynamical systems}
\label{maxlikelihood}

\paragraph{Setup.}
To formulate a cell-specific  model, we consider each cardiomyocyte as
a spatially localised system $C$ described by a set of ordinary
differential equations of the form
\begin{gather}
  \label{eq:odes}
  \frac{d}{dt} {x} = f(t, x; \vartheta), ~~~ x(0)=x_0, ~~~ y = s(x).
\end{gather}
\looseness=-1
Here  $t\in \mathbb{R}^1$ denotes time, $x(t,\vartheta,x_0) \in
\mathbb{R}^d$ is a vector of state variables with initial values
$x_0\in \mathbb{R}^d$, and ${\vartheta}\in \mathbb{R}^k$ is a vector of
model parameters, $y(t,\vartheta,x_0) \in \mathbb{R}^n$ is a vector of
observable outputs and $f$ and $s$ are functional relationships, often
called ``the model'' in this context.
We consider a dataset
\begin{gather}
D:=\Big\{(t_j,Y_j)\Big\}_{j = 1}^{K}.
\label{eq:dataset}
\end{gather}
of $K$ experimental values $Y_j\in\mathbb{R}^n$ of the
observables $y$ measured at discrete times $t_j$.

\looseness=-1
\paragraph{Noise model.}
Measurements are subject to random observational errors and
inherent fluctuations within the complex system $C$. 
Thus, we assume that the experimental data values $Y_j$ are realisations of
normally distributed independent random variables $\mathsf{Y}_j$ with mean values
$E[\mathsf{Y}_j]=y(t_j,\theta,x_0)$ and unknown but identical
variances  $\sigma^2=\sigma_j^2$, id est
\begin{gather}
\label{eq:gauss}
Y_j \sim \mathcal{N}\big(y(t_j,\theta,x_0),\sigma^2\big).
\end{gather}
These commonly made assumptions are justified by the universality
of the Gaussian probability distribution $\mathcal{N}$.
There is no essential distinction between the unknown variance parameter $\sigma^2$, the
initial values $x_0$ and the model parameters $\vartheta$ and we group 
them in the vector $\theta \subseteq (\vartheta,x_0,\sigma) \in
\mathbb {R}^l$ with $l\le d+k+1$. 
We now turn to their estimation.

\paragraph{Maximum likelihood.}
We invoke the maximum likelihood principle to (i) find point estimates
$\hat\theta$ of these parameters, (ii) measure the standard
errors $\sigma_{\hat\theta}$ of the estimates, and (iii) quantify the
goodness of fit. 
Under our assumptions, the likelihood function of the data $D$, defined
as the joint probability density $P$ to measure
$Y=\{Y_j, j=1\dots K\}$ considered as a function of the parameters
$\theta$,
is
\begin{align}
\label{eq:likelihood}
L(Y;\theta) =& P(Y|\theta) = \prod_{j=1}^K P(Y_j|\theta) = \prod_{j=1}^K 
\frac{1}{\sqrt{2\pi\sigma^2}}
\exp\left(-\frac{1}{2} \frac{\big(Y_j-y(t_j,\vartheta,x_0)\big)^2}{\sigma^2}\right). 
\end{align}

\paragraph{Point estimates.}
Proceeding from the definition of likelihood, the maximum-likelihood principle postulates that
the best point estimate $\hat\theta$ of the parameter values is given
by the values of $\theta$ for which $L(Y; \theta)$ attains its global
maximum
\begin{gather}
\hat\theta = \arg\max\limits_{\theta\in\Theta} L(Y;\theta),
\label{eq:argmax}
\end{gather}
where $\Theta\subseteq \mathbb{R}^l$ is an appropriately constrained
region of the parameter space. 
The evaluation of the maximum-likelihood estimator \eqref{eq:argmax}
now becomes a mathematical optimisation problem that may be solved
by numerous methods \citep{Aragon2019}. 

\paragraph{Errors of estimation.}
We assume that errors in numerical optimisation are negligible in
comparison to errors in the estimation of $\theta$ that arise from
experimental noise. Thus, we seek the standard
error of the estimates as the square root of the variance
$\text{Var}[\hat\theta(Y)]$. This, in turn, can be related to the
variance $\hat\sigma^2$ of the voltage measurements, \rev{which can
equivalently be interpreted as the standard error of voltage
estimation,} and  found as a component of \eqref{eq:argmax}.
Approximating the data $Y$ by the model $y$ and then the relationship $\hat\theta(y)$
by the linear terms of its Taylor expansion about the expectation
$E[y]$, we find that standard errors of the estimates can be measured by
\begin{align}
\label{eq:stderr}
  \sigma_{\hat\theta} = \hat\sigma \sqrt{\text{diag}\big([J^T J]^{-1}\big)}, ~~~~~ J=
\nabla_\theta y\big|_{\hat\theta},
\end{align}
where $\nabla_\theta$ denotes the gradient of partial derivatives with
respect to $\theta$ and $T$ denotes a matrix transpose.
Here, we have taken advantage of the inverse function theorem and
expressed the parameter covariance matrix in terms of the Jacobian
matrix $J$ of the inverse relation $y(\theta)$ as this is more easily
evaluated by a minor extension of problem \eqref{eq:odes}.

\paragraph{Goodness-of-fit.}
To assess the goodness-of-fit we exploit the fact
that if $\mathsf{Y}_j$ are normally distributed as assumed,
and if $y(t_j,\theta)$ is linear in $\theta$, then the
sampling distribution of the sum of squared errors 
\begin{gather}
\label{eq:chi2}
  \hat\chi^2(\hat\theta) =\sum_{j=1}^{K}
  \frac{\big(Y_j-y(t_j,\hat\vartheta,\hat x_0)\big)^2}{\hat\sigma^2},
\end{gather}
must be a chi-squared distribution with $\nu=K-l$ degrees of freedom,
see e.g.~\citep{Riley2006}.
Specifically, let $H_0$ be the null hypothesis asserting assumptions
are correct, and the quality of fit is good. To test $H_0$, we
calculate the probability $p$ under $H_0$ of obtaining a value $\chi_{\nu}^2$
that
is larger than the value $\hat\chi^2(\hat\theta)$ measured
at the best parameter estimates $\hat\theta$,
\begin{gather}
  \label{eq:pvalue}
p= \text{Pr}\Big(\chi_\nu^2  \ge \hat\chi^2(\hat\theta)\,\Big|\, H_0 \Big) =
\int_{\hat\chi^2(\hat\theta)}^\infty P(\chi_{\nu}^2)\, d\chi_{\nu}^2,
\end{gather}
where $P(\chi_\nu^2)$ is the probability density function of the
chi-squared distribution with $\nu$ degrees of freedom.
We reject $H_0$ at significance level $\gamma$ if $p<\gamma$. 
Linearity of $y(t;\theta)$ is not satisfied by \eqref{eq:odes}, but
classical texts \citep{Press2007} advise that this test is also
acceptable in non-linear cases.

\paragraph{A population of isolated cells.}
Finally, to extend the analysis to population level, we 
consider a set of $N$ cardiomyocytes with associated experimental
data  
\begin{gather}
\mathcal{C}=\big\{ C_i\big\}_{i=1}^N,
  ~~~~ \mathcal{D}=\big\{D_i\big\}_{i=1}^N.
\end{gather}
We find parameter estimates \eqref{eq:argmax}, their standard
errors \eqref{eq:stderr}, and evaluate the test \eqref{eq:pvalue} for
each one.
The population of system-specific models consists of all accepted fits
\begin{gather}
  \mathcal{M}=\big\{ \hat\theta_i \pm \sigma_{\hat\theta,i}\big\}_{i=1}^N,
  ~~~  \text{s.t.} ~~~ p > \gamma.
  \label{eq:pop}
\end{gather}

\subsection{Fluorescence recordings of action potential waveforms}
\label{sec:experiment}
\looseness=-1
We fit newly recorded action potential waveforms from
1228 rabbit ventricular myocytes. Cardiomyocytes were obtained from
eight male New Zealand White rabbits.
Enzymatic
cardiomyocyte isolation and fluorescence-based recording of
transmembrane voltage from individual cardiomyocytes were performed as
previously described in \citep{Lachaud2022}. Myocytes isolated from the
free wall of the left ventricle were loaded (16 minutes, room
temperature) with 0.08 $\mu$L/mL FluoVolt (Thermo Fisher Scientific). For
recordings, myocytes were bathed in Krebs-Henseleit solution
containing (in mM): 120~NaCl, 1.8~CaCl$_2$, 20~HEPES, 5.4~KCl,
0.52~NaH$_2$PO$_4$, 3.5~MgCl$_2\cdot$6H$_2$O, 20~taurine, 10~creatine
and 11~glucose (pH~7.4 at 37$^\circ$C).
Myocytes were subjected to field stimulation with 40~V,
2~ms pulses at a frequency of 2 Hz.
Cells were stimulated for 5~min before recordings. Fluorescence
intensity was then measured at 10~kHz frequency for
a further 2.5~s, yielding a train of 5 action potentials. These were then
temporally averaged to provide a single waveform for each cell.
\rev{The beat-to-beat variability was monitored and assessed as described previously
in \citep{Lachaud2022} and was less than 2\% for measurements from
individual myocytes.}
All recordings were made at 37$^\circ$C.
Cells can be split into sub-populations from distinct apical/basal
and  endo/mid/epicardial sub-regions and from different animals.
However, here we regard all measured cells as a single large
myocyte population $\mathcal{C}$ with associated experimental data
$\mathcal{D}$, consisting of individual time series $D_i$ of
fluorescence intensities $\mathcal{V}_{i,j}$ measured at
times $t_j$,
\begin{gather}
  \label{eq:data}
\mathcal{D} = \left\{ D_i = \Big\{(t_j=j\Delta t,
\mathcal{V}_{i,j}), ~ \Delta t=10^{-4}
\text{s}\Big\}_{j=1}^{K}\right\}_{i=1}^N, ~~~~~ K=5000, ~ N=1228.
\end{gather}
Examples of nine single averaged action potential wafevorms are
shown in Figure \ref{fig010} and those of all accepted biological
cells are included in Supplementary Figure \ref{supfig010}. 

\subsection{Baseline action potential model}
\label{sec:baseline}

\paragraph{Model description.}
\looseness=-1
The \textcite{Shannon2004} model of the
rabbit ventricular myocyte is used as a baseline model for the
experimental data $\mathcal{D}$. 
In this model, cells are represented as a collection of four
compartments: sarcoplasmic reticulum, junctional cleft,
subsarcolemmal space and cytosolic bulk. 
Ions of species Ca$^{2+}$, Na$^+$, K$^+$, and Cl$^-$ are exchanged between 
compartments, and with the extracellular space by facilitated
diffusion and active transport.
The model characterises the instantaneous state of a cell by a
set of 38 state variables including ion channel gating variables, 
ryanodine receptor variables, and concentrations of each of
the four ionic species in free and bound states and for each of the four
compartments. Here, these variables are denoted by a vector 
$z\in\mathbb{R}^{38}$ and obey a set of nonlinear ordinary
differential equations with rates given by a vector field $g$.
The ion transport fluxes give rise to 15 electric currents, namely
a fast Na$^+$ current $I_\text{Na   }$,
a L-type Ca$^{2+}$ current $I_\text{CaL  }$,
rapid and slow components of the delayed rectifier K$^+$ current
$I_\text{Kr   }$ and  $I_\text{Ks   }$, respectively,
inward rectifier K$^+$ current $I_\text{K1   }$,
fast and slow transient outward K$^+$ currents $I_\text{tof  }$ and $I_\text{tos  }$, respectively,
Ca$^{2+}$-activated Cl$^-$ current $I_\text{ClCa}$,
Na$^+$/Ca$^{2+}$ exchanger current $I_\text{NaCa }$,
Na$^+$/K$^+$ pump current $I_\text{NaK  }$,
sarcolemmal Ca$^{2+}$ pump current $I_\text{Cap  }$,
and background Na$^+$, K$^+$, Ca$^{2+}$, and Cl$^-$ currents
$I_\text{Nab  }$, $I_\text{Kp }$ $I_\text{Cab  }$, $I_\text{Clb  }$, respectively,
denoted by $I_k, k=1\ldots15$ below.
The ion currents are simultaneously modulated by and drive changes in
the voltage $V$ across the sarcolemma which, in turn, is modelled as
a circuit with a capacitor and a resistor connected in parallel.
In the experiments cells are excited by an additional
stimulus current, $I_\text{stim}$, in the form of a
train of rectangular pulses.
With these notations, the model of \textcite{Shannon2004}
can be specified by setting in equations \eqref{eq:odes}
\begin{align}
  \label{eq:shannon}
  x = [V, z]^T,  ~~~~~ y = s(x): =V, ~~~~~
  f = \Big[\sum_{k=1}^{15} I_k(V,z;\vartheta) +
    I_\text{stim}(t,\vartheta), ~~g(V, z; \vartheta)\Big]^T, 
\end{align}
where 
$\vartheta\in\mathbb{R}^{140}$ is a vector of 140 model parameters
including the stimulus duration, amplitude and frequency.
Initial values of all state variables must be included, extending the vector
of parameters to 179, i.e. $[\vartheta,x_0]^T \in \mathbb{R}^{179}$. Specific
parameter values 
and algebraic expressions for the currents and the nonlinear functions
$f$ and $g$ are provided in the original
publication \citep{Shannon2004}, and we use 
an error-free machine readable implementation
from the CellML model repository \citep{Lloyd2008}.
The only modification made is that the reversal potential of sodium
ions across the sarcolemma is changed from a Nernst equation to the fixed value
$ E_\text{Na,SL}=-15  ~\text{mV} $
to mimic experimental waveforms $D_i$ where 
the ``spikes'' of the peak voltages are systematically observed  to be
smaller than those of the original Shannon model.
This implementation reflects the uncertainty as to the electrical
conditions that initiate the action potential and the model's ability
to reproduce aspects of field stimulation and explains why this study
is focussed on fitting the secondary repolarisation phase. 

\paragraph{Numerical solution.}
The model solution $x(t; \vartheta, x_0)$ and observables $y(t,\theta,
x_0)$ are required to evaluate the likelihood
function \eqref{eq:likelihood} at given parameter values.
Solutions are obtained numerically using the CVODES
method 
from the {SUNDIALS} suite of nonlinear and differential equation solvers \citep{hindmarsh2005}
with absolute and relative tolerance settings of $10^{-6}$ and
$10^{-8}$, respectively.
The Myokit ``interface to cardiac cellular
electrophysiology'' \citep{Clerx2016} is used to access CVODES.

\subsection{Estimands and optimisation details}
\label{sec:OptimisationDetails}
\paragraph{Estimands.}
It is computationally infeasible to include all 179 parameters of
the Shannon model in the parameter estimation
problem \eqref{eq:argmax}.
In this study we keep the vast number of
parameters fixed to their original values and seek to calibrate only
the following eight model parameters and the standard variation of noise
\begin{gather}
\label{eq:params}
  \theta = [
    \overline{G}_\text{Kr},
    \overline{G}_\text{Ks},
    \overline{G}_{K1},
    \overline{G}_\text{tos},
    \overline{G}_\text{CaL},
    \overline{G}_\text{Clb},
    \overline{I}_\text{NaK},
    \overline{I}_\text{NaCa}
, \sigma]^T \in \mathbb{R}^9.
\end{gather}
\revii{These eight parameters were chosen following our earlier
local sensitivity analysis of this model published in \citep{Lachaud2022}.}
Here subscripts denote ion currents as introduced in
subsection \ref{sec:baseline}, with $\overline{G}$ being the maximal
conductance and $\overline{I}$ being the maximal density of
currents with Ohmic and with Goldman-Hodgkin-Katz mathematical
formulations, respectively. 
In the following, results are quoted as a proportion
$\alpha\in\mathbb{R}^9$ of the published baseline  values
$\check{\theta}$ and the estimates of the actual cell-specific
parameter values can be obtained by taking the Hadamard element-wise
product $\theta = \alpha \odot \check{\theta}.$
\rev{The proportion factors $\alpha_k$ represent ``relative strengths'' of
currents compared to the baseline, as used for instance in \citep{Lawson2018}.} 
The factors $\alpha_k$ are assumed positive in order to preserve both 
model dynamics and the physiological interpretation of the calibrated
parameters.

\paragraph{Optimisation details.}
The global maximisation \eqref{eq:argmax} \rev{of the logarithm of the
likelihood function \eqref{eq:likelihood}} is performed using the
Covariance Matrix Adaptation Evolution Strategy algorithm (CMA-ES)
of \citep{Hansen2006} as implemented in the Python module 
PINTS \citep{Clerx2019}.
The CMA-ES is a gradient-free method \rev{designed for high-dimensional,
ill-conditioned, non-convex problems.
It initiates a population of candidate estimates by sampling from a
multivariate normal distribution with a mean and a covariance matrix
given by an initial guess. It evaluates the log-likelihood of these
candidates by simulating the Shannon model \eqref{eq:shannon} and comparing
the computed and the measured voltage traces as per equation \eqref{eq:likelihood}. It then selects a
sub-population of the most likely candidates and uses them to
compute an updated mean and covariance matrix, thus effectively finding the
direction of higher likelihood. The steps of sample generation,
selection and update are then repeated until given convergence
criteria are met.}
A population of 96 initial candidates is used with \rev{means
centred at unity corresponding to the published Shannon model baseline
values and with diagonal covariance matrices with variances set to
16.66. The variance values represent one-sixth of the boundary ranges
for the parameter search which were taken as $10^{-4}$ to $10^2$ for all
parameters}. The optimisation is terminated when estimates exhibit a
relative change of less than $10^{-6}$ over the last 100 iterations of
the algorithm.    

\section{Results and discussion}
\label{sec03}

We now proceed to describe the population of cell-specific
action potential models obtained by fitting the
\textcite{Shannon2004} model to voltage-sensitive fluorescence measurements in rabbit ventricular
myocytes.

\subsection{Illustrative demonstration}
\label{sec:illustrate}
\begin{figure*}[t]
\hspace*{-5mm}
\begin{overpic}[width=1.05\textwidth]{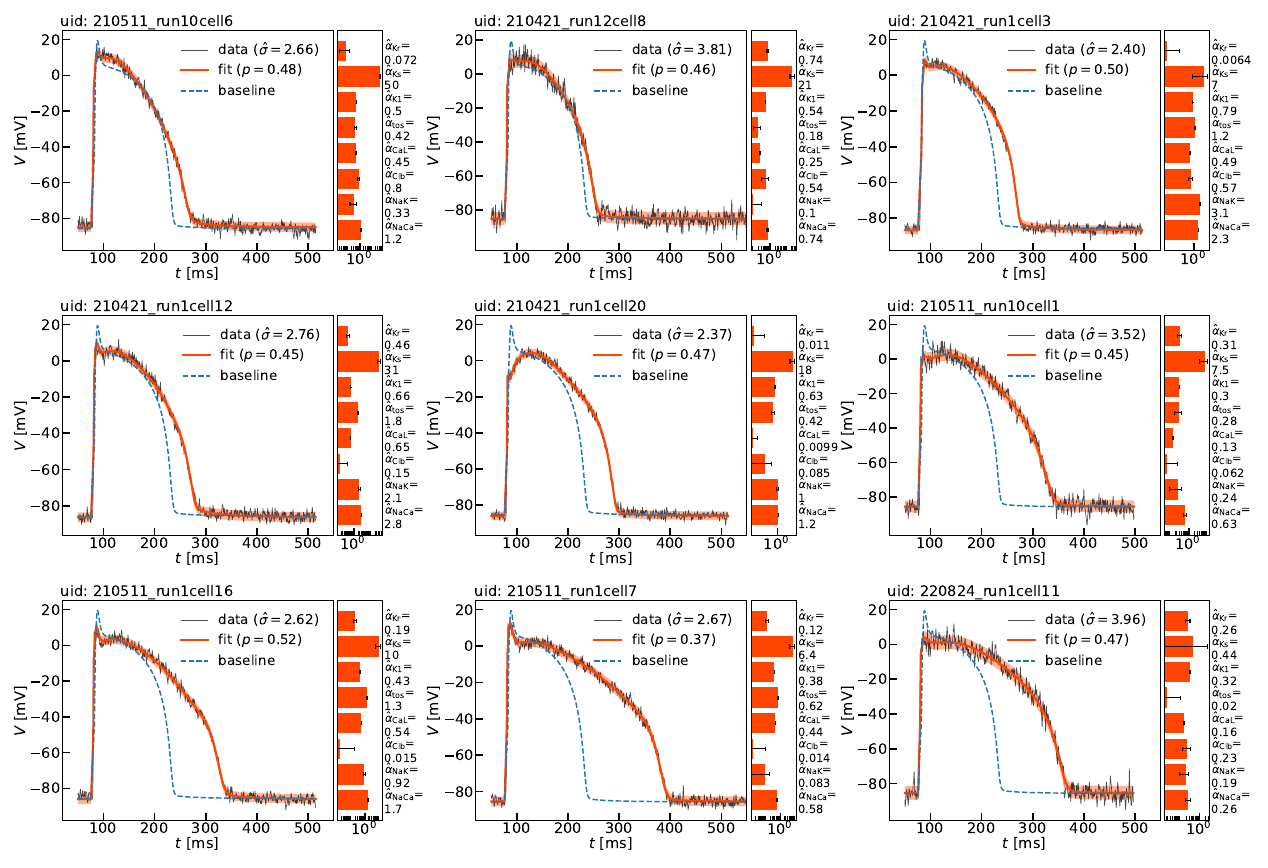}
\end{overpic} 
\caption{Examples of parameter estimations for nine typical biological
myocytes. Cells are identified by unique identifiers (uid). The
experimentally-measured action potential waveforms $\{(t_j,
\mathcal{V}_j), j=1,\cdots,5000\}$ are shown by thin 
black lines. Action potential waveforms $\hat V(t)$ computed from the
accepted fits of the Shannon model \eqref{eq:shannon} are shown by
thick orange-red lines. 
Estimates of the standard deviation $\hat\sigma$ of noise in voltage
measurements are quoted in the legends, and are \rev{illustrated by the width of
the semi-transparent orange-red strips centred on $\hat V(t)$}. The
corresponding point estimates $\hat \alpha$ of the parameter values
and their standard errors $\sigma_{\hat\alpha}$ found
by \eqref{eq:argmax} and \eqref{eq:stderr}, respectively, are quoted
and illustrated in a separate bar chart for each cell. The 
goodness-of-fit values $p$ measured by \eqref{eq:pvalue} are listed in the legends. The action
potential waveform $\check{V}(t)$ of the baseline Shannon model is
shown by blue dashed lines for comparison. 
\label{fig010}
} 
\end{figure*}

To illustrate the parameter inference procedure on specific examples,
we discuss first a small subset of nine rabbit ventricular myocytes.
Figure  \ref{fig010} shows visualisations of (a)
the available single-cell experimental data, (b) the accepted fits
with their goodness-of-fit measures, (c) the inferred model parameter
estimates with their standard errors, along with (d) a direct
comparison to the baseline Shannon model for  these nine typical
myocytes. Full details for the remaining 1180 accepted  cells are
included in Supplementary Figure \ref{supfig010} and Supplementary
Tables \ref{tab:estimates} and \ref{tab:stderr}.

\paragraph{Features of experimental waveforms.}
In common with the action potentials of all excitable cells, ventricular
action potentials are large transient excursions away from electric
potential equilibrium that exists across sarcolemmas \citep{Amuzescu2021}.
A good example of the generic morphology of the ventricular action
potential is provided by the voltage component $\check{V}(t)$ of the
solution to the baseline  Shannon model plotted in Figure \ref{fig010}.
The experimentally measured waveforms $\mathcal{V}_j(t)$ exhibit similar behaviour but show several
distinctive features as seen in Figure \ref{fig010}.
In particular, we note that the experimental action potentials $\mathcal{V}_j(t)$
plotted in the figure are longer than that of the baseline model
$\check{V}(t)$ as measured by their  action potential duration
(APD$_{90}$). This is also true for the majority of all 1180
biological cells. The duration APD$_{90}$ is defined as 
the time interval between depolarisation upstroke and repolarisation
downstoke measured at 90\% of the waveform amplitude; durations
APD$_{x}$ can be defined similarly. 
Another characteristic feature of the measured traces $\mathcal{V}_j(t)$
is the absence of significant spikes, and rather weak notches
afterwards. Some experimental waveforms feature no spikes at all as
seen, for example, in cell {\sffamily uid: 210421\_run1cell20}. This
can be understood as a case 
of a fast sub-threshold but slow over-threshold response to the excitation
stimulus \citep{Biktashev-2008}, where for the
particular cell the stimulus current has insufficient amplitude or
duration to trigger a response in the fast sodium current
$I_\text{Na,SL}$ but is large enough to trigger a response in other
slower currents so that the voltage transitions to the quasistable
manifold of the plateau ``from below''. It is the characteristic absence of
pronounced spikes in experimental waveforms that lead us to reduce
the baseline value of the reversal potential  $E_\text{Na,SL}$ of the
sarcolemmal sodium current to $-15$ mV, as mentioned in
section \ref{sec:baseline}. All other effects of this change on the
baseline model are negligible \rev{as demonstrated in Supplementary Figure \ref{revsupfig:ENaSLcheck}.}

\paragraph{Effect of noise.} 
Random noise appears to have different signal-to-noise
ratio in different cells as seen  in the examples of Figure \ref{fig010}.
The random noise has a characteristic time scale shorter than the
time scales along the slow pieces of the action potential trajectory,
e.g.{} slower than the time scale of evolution on the plateau and the
resting potential, but comparable to the fast time scale of
the upstroke. This makes the noise transversal to 
the slower pieces of the action potential resulting in difficulties in 
ascertaining exact values of voltage along the plateau and at
resting equilibrium without resorting to long temporal running averages.
It also makes it difficult to recognise waveform features that are
comparable in time scale and amplitude to noise as it is impossible to
time-average there. The spikes, and to a lesser
extent the notches, are the main such morphological features and, as a
result of noise, the peak voltage $\mathcal{V}_\text{peak}$ cannot be
accurately determined. 
On the other hand, processes on faster time scales and with larger amplitudes
than noise, e.g.{} the upstroke and the action potential
duration, are less affected by noise and can be determined relatively
accurately.

\rev{Supplementary Figure \ref{revsupfig_hist_autocorr} illustrates
further the properties of experimental noise. Sub-plot (a)  provides a histogram of
residual differences between the true experimental values of the
voltage and the model estimated values of the voltage for all moments in time,
$\{ e_i = \mathcal{V}_j - \hat V(t_j)$, $i = 1, \cdots ,5000 \}$. The
mean value of the histogram sample deviates insignificantly from zero,
which is due to the finite size of the sample (5000). The standard deviation of
the sample is identical to the standard deviation (standard error of
estimation of voltage) found by optimisation. A Gaussian
distribution with these parameter values closely captures the shape
of the histogram as shown in the Figure. This demonstrates that the
assumption of normality of errors is well satisfied.}
\rev{Sub-plot (b) shows the autocorrelation
(Pearson's correlation coefficient $r$) between voltage residual values
at moments $t$ and $t+\text{lag}\Delta t$.}
\correct{Non-negligible correlation is observed with the preceding 10 to
15 values which suggests that the assumption of independence and
identical distribution of errors is not well satisfied. A more
general autoregressive (integrated) moving-average noise model may
have been more accurate to use. Inevitably, this involves estimating a
larger number of parameters and will be left for future refinements.}

\rev{\paragraph{Fluorescence to voltage mapping.}
Voltage-sensitive fluorescence measurements do not provide absolute
voltage values. To convert fluorescence intensity to voltage
we have chosen to map the time-averaged fluorescence intensity at the
plateau to $0$ mV and the time-averaged fluorescence intensity at rest
to $-86$ mV with linear scaling between. The plateau is chosen as noise prevents accurate capturing of the signal
spike as discussed above. The values $0$ mV and $-86$ mV
are the values of $\check{V}_\text{plateau}$ and
$\check{V}_\text{rest}$ of the voltage in the baseline Shannon model, respectively. The
examples of Figure \ref{fig010} demonstrate that this works well.
}

\paragraph{Model fits.}
The accepted model fits for the nine cells of our illustrative subset
are shown  Figure \ref{fig010}.
The estimated parameter values ${\hat\theta}$
found by maximising the likelihood function via \eqref{eq:argmax} are
explicitly stated as proportions $\hat \alpha$ relative to the
baseline values $\check{\theta}$ and illustrated by bar charts for each cell.

The standard errors of these estimates, denoted as
$\sigma_{\hat\alpha}$, were determined using
equations \eqref{eq:stderr} and are depicted as error bars overlaid on
the bar-chart values in Figure \ref{fig010}.
Since numerical errors are assumed negligible, these standard errors are 
measures of uncertainty in the estimation process. Uncertainty is
due to random noise and to model selection choice. The latter
manifests itself here as alternative possibilities for selection of
the number of parameters to be estimated within the chosen baseline
model \citep{Shannon2004}.
It is pleasing to find that, on average, six out of the eight
estimates are obtained with small uncertainty for the cells in the
illustrative subset.
However, a few parameter estimates consistently exhibit large standard
errors across all fits, and this uncertainty varies across the estimands
for different cells. This issue is further explored in the subsequent
subsection. 

The synthetic action potential waveforms computed from the Shannon
model using the quoted cell-specific parameter estimates are superimposed onto
the experimental waveforms in Figure \ref{fig010}, displaying a close
match across all cells and action potential phases. 
The excellent visual goodness of fit is substantiated by evaluating
the formal measure \eqref{eq:pvalue}. Values of $p$ are given in the
Figure and are close to 0.5 for all nine cells. 
Recall that $p$ represents the probability of obtaining a weighted sum
of squared errors $\chi_\nu^2$ larger than the value
$\chi_\nu^2(\hat\theta)$ actually measured in a final fit
under the null hypothesis that the assumed model is correct. Since the
objective is to minimise $\chi_\nu^2$, large values of $p$ are considered
good fits, see discussion in Ch 15  of \citep{Press2007}. 

\rev{The standard deviation $\hat \sigma$ of the experimental signal from
the synthetic mean is a measure of experimental noise. This quantity is 
estimated as a component of the parameter vector during log-likelihood
maximisation as mentioned above. The estimated values of $\hat \sigma$
for the nine cells in our illustrative subset are visualised by
semi-transparent strips of width $2\hat\sigma$ centred on the model
synthetic waveforms in Figure \ref{fig010}. When superimposed 
onto the noisy experimental traces these values and the corresponding
strips they define appear to capture noise levels very well.
}

Finally, we mention that maximum likelihood estimation takes approximately 30
minutes of wall clock time when running a parallel implementation of the
CMA-ES method with 96 threads on a small multi-user computing server.
Exact computation times depend on the data being fitted, proximity of the initial
guess to maximum likelihood, the random nature of the CMA-ES method
itself, but typically convergence is achieved in 300 to 500 generations
of CMA-ES method of which 100 have absolute change smaller than
$10^{-6}$ as required by the imposed termination criterion.

\begin{figure*}[t]
\hspace*{-6mm}
\begin{overpic}[width=1.05\textwidth]{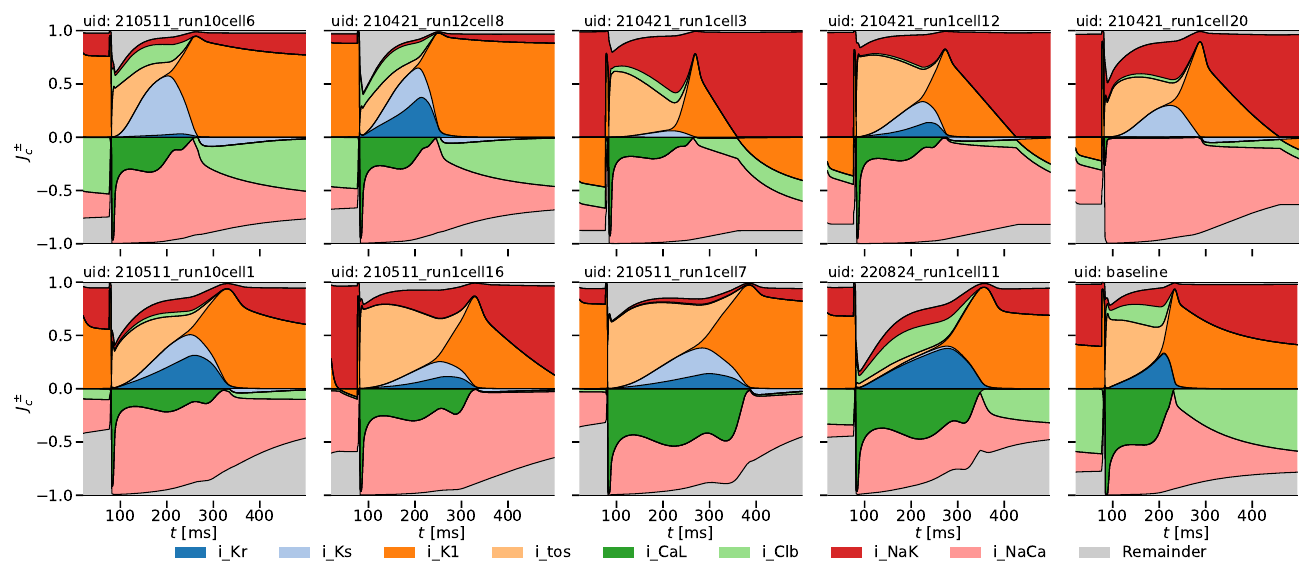}
\end{overpic} 
\caption{\revii{Normalised positive and negative
cumulative currents $J^{\pm}_c$ as
functions of time computed from the
cell-specific models of the  nine myocytes shown in Figure
\ref{fig010}.
The cumulative currents are defined by
equations \eqref{eq:cumcurrs} and coloured as specified in the figure
legend. The cumulative currents of the Shannon baseline model are
also for 
comparison.}
\label{fig020}}
\end{figure*}

\paragraph{Model analysis and prediction.}
The chief purpose of a mathematical model is to conceptualise a
real-world system and to enable its formal analysis and forward
prediction. To illustrate this trivial remark,
we plot in Figure \ref{fig020} normalised positive and negative
cumulative currents $J^{\pm}_c$ for the nine cells
illustrated in Figure \ref{fig010}. These quantities are defined as 
\begin{subequations}
\label{eq:cumcurrs}
\begin{align}
   \left\{J^{\pm}_c (t) = \left.\pm\sum_{k=1}^{c}
    I_k^{\pm}(t)\right/\sum_{k=1}^{K=15}
    I_k^{\pm}(t)\right\}_{c=1}^{8}, 
     ~~~~~~~~~~~
    I^{+}_k(t)=\max\big(I_k(t),0\big), ~~~
    I^{-}_k(t)=\min\big(I_k(t),0\big), 
\end{align}
where  $I_k$ are the \textcite{Shannon2004} model currents indexed by the
elements of the ordered set
\begin{gather}
[k]_1^{15} = [\text{Kr, Ks, K1, tos, CaL, Clb, NaK, NaCa, Na, Nab,
        tof,  ClCa,  Cab, Cap, Kp}].
\label{eq:indices}
\end{gather}
\end{subequations}
The cumulative currents preserve their strict ordering for all time, i.e.
$|J^\pm_c(t)| < |J^\pm_{c+1}(t)|$,  $c=1\cdots 7$.
When cumulative currents are plotted in reversed
order from the largest in the background to the smallest in the
foreground and as functions of time, complement regions between the
curves of $J^\pm_c(t)$ and 
$J^\pm_{c+1}(t)$ represent the contribution of Shannon current
$I_{c+1}(t)$ added to the contribution of the preceding $c$ currents.
Thus the cumulative currents are an equivalent representation of the
physiological ionic currents in the model with the advantage
that they can be conveniently plotted one over another, the positive and negative parts
of currents can be plotted separately and the relative contribution
of each current to the total can be shown. For comparison, the
cumulative currents corresponding to the baseline Shannon model are
shown in a ``mirror image'' against each myocyte to illustrate
the predicted differences in internal ion dynamics of each
cell. We point to the fact that the cumulative contribution of
those  currents that have been kept fixed during parameter
estimation is small. This cumulative ``remainder'' current is plotted
in grey in Figure \ref{fig020}.
\rev{The actual ionic currents in the selection of models are
shown in Supplement Figure \ref{revsupfig01}. We note in Figure \ref{fig020}
that the background chloride current near the resting state can be as
low 0\% of the total for some cells or as high as 90\% of the total
for others. However, Supplementary Figure \ref{revsupfig01} shows that
the total current at rest is very nearly zero.}
Since ionic currents have not been measured in our experiments, this
calculation presents an example of model forward prediction.  
Similarly to the example of Figure \ref{fig020}, cell-specific
Shannon models can be used, with equal ease, to derive and compute
additional quantities that are impossible to, or have not been,
experimentally measured. For instance, time series of ionic
concentrations can be easily calculated, and cellular responses under
different conditions including drug treatments can be readily
predicted. Detailed applications of such type will be considered in future work.

The features and the overall quality of parameter estimation,
illustrated here on the nine example myocytes, are typical for the
entire population of 1228 fitted cells, as evidenced in the
Supplementary Material. 

\begin{figure*}[t]
\begin{center}
\hspace*{-5mm}
\begin{overpic}[width=0.95\textwidth]{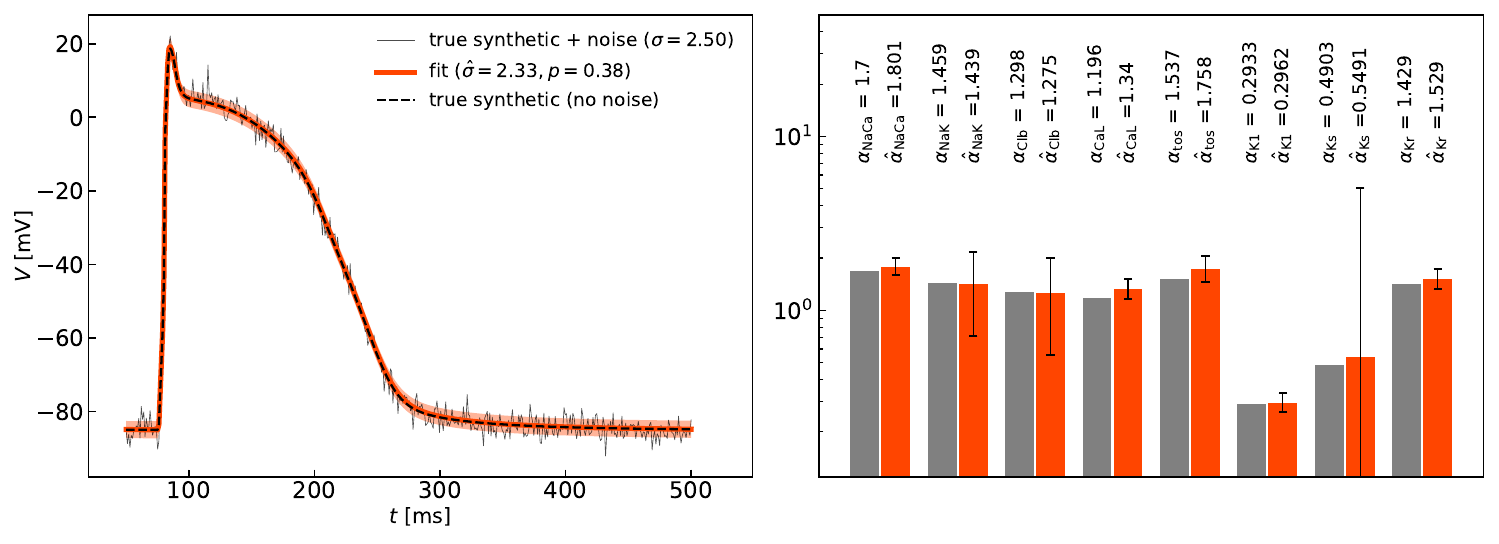}
  \put(1,36) {{(a)}}
  \put(52,36) {{(b)}}           
\end{overpic}
\end{center}
\caption{\rev{Fitting to synthetic data. Given randomly selected, but known,
parameter values (grey barchart in panel (b)) a synthetic action potential
trace was generated (black dashed curve in panel (a)). Gaussian random noise
with mean 0 and standard deviation $\sigma = 2.5$ was added to values
of $V_i$ to produce a synthetic noisy action potential
trace (thin black curve in (a)). The noisy signal was refitted as
described to produce a fitted AP trace (thick red curve in
(a)). The red barchart in (b) shows the point estimates of the
parameter values with standard errors shown in black error bars
with end caps. }
\label{fig030}}
\end{figure*}

\subsection{Synthetic data test}
\label{sec:synthetic}

To further validate the inference methodology we performed parameter
fitting of synthetic data as illustrated in \ref{fig030}. The test allows evaluation
of absolute errors which are not available when fitting experimental
data.

\paragraph{Synthetic data generation.}
We drew random values for each of the
eight parameter estimands $\alpha_k$ of equations \eqref{eq:params} from a
continuous uniform distribution $U\big([0.1, 2]\big)$.  The specific
values are given in the right panel of Figure \ref{fig030}.
With these we computed a
numerical solution of the Shannon model \eqref{eq:shannon} to generate
a ``synthetic'' clean action potential waveform $V(t)$. We added to this signal
normally distributed random noise with standard deviation $\sigma =
2.5$ to obtain a synthetic noisy action potential $V(t)+v(t)$, where
$v(t)\sim \mathcal{N}(0,\sigma^2)$.
There was no specific consideration in
choosing the interval of the uniform distribution to be $[0.1, 2]$ other
than to avoid ``non-action potential'' solutions which often occur
when parameter values are taken far from the baseline values of
unity. The value of the standard deviation was selected to be similar
to the ones found in the illustrative examples shown in
Figure \ref{fig010}. The resulting synthetic action potential trace  
without and with noise is plotted in Figure \ref{fig030} and the
noisy one is visually indistinguishable from typical measurements as
seen in Figure \ref{fig010} and in Supplementary Figure \ref{supfig010}.

\paragraph{Quality of fit.}
The synthetic noisy action potential  was then refitted following steps identical to those
involved in the parameter estimation for biological
myocytes. The results of the test are presented in Figure \ref{fig030}.
We find absolute error in estimating the standard deviation of noise to be
$|\sigma - \hat\sigma| = 0.17$ corresponding to a relative error 
$|\sigma - \hat\sigma|/\sigma=0.068$. We find similarly small relative errors of the order
$O(10^{-2})$ for the values of all estimands as illustrated in the
right panel of Figure \ref{fig030} where the estimated parameter
values and the absolute errors between estimates and true values of
the estimands $|\alpha - \hat\alpha|$ are given.
These absolute errors are well within, in fact significantly
smaller than, the standard errors of estimation determined by
equation \eqref{eq:stderr}. It must be noted that, in contrast,
the standard errors of estimation $\sigma_{\hat\alpha_k}$  are particularly large for some
estimates. In this instance $\hat\alpha_\text{Ks}$, and to a lesser degree 
$\hat\alpha_\text{Clb}$ and $\hat\alpha_\text{NaK}$, have large
standard errors even though their values are very accurate estimates
of the true ones. We interpret this as an
indication that the model solution is not very sensitive to the values
of these specific parameters. 
Naturally, sensitivity varies across the eight-dimensional parameter
space and model solution may be 
more or less sensitive or insensitive to different estimands. This can be observed in
myocytes {\sffamily uid: 210421\_run1cell3} and {\sffamily uid:
  220824\_run1cell11} of Figure \ref{fig010} where $\hat\alpha_\text{Kr}$
and $\hat\alpha_\text{tos}$ show large standard errors of estimation
indicating that the model is less-sensitive to these quantities in these
instances. However, the small relative errors
$|(\alpha_k-\hat\alpha_k)/\alpha_k|$ found in the synthetic test give
confidence that estimates are accurate even when parameter uncertainly
as measured by $\sigma_{\hat\alpha_k}$ is significant.

The most important test of the quality of the fit is, of course, the
agreement between the synthetic clean action potential trace $V(t)$
and the trace $\hat V(t)$ computed from the refitted model using the estimated
parameter values. The two traces visually overlap as seen in the left
panel of Figure \ref{fig030}. To quantify the difference between them
precisely, we computed the 
values of the absolute and the relative root-mean square
averages of the errors between the two
$$
\bar {e}_V =  1.0\times10^{-2} \text{mV}, ~~~~ {\bar {e}_V}\big/{V_\text{amp}} = 1.0\times10^{-4},    ~~~
(V_\text{amp} = 103.78 \text{mV}).
$$
These are defined as usual by
$$
\bar {e}_V =  \frac{1}{N}\sum_{i=1}^{N=5000} \sqrt{\big(V(t_j) - \hat
  V(t_j)\big)^2}, ~~~~~ V_\text{amp} = \max_t V(t) - \min_t V(t).
$$

In summary, the small relative errors from the known true values in
this synthetic test demonstrate the excellent quality of the fit shown
in Figure \ref{fig030}.
While true errors are unknown for biological cells, we believe that
fits are of a similarly good quality for all cells. 

\begin{figure*}[t]
\hspace*{-6.5mm}
\begin{overpic}[width=1.05\textwidth]{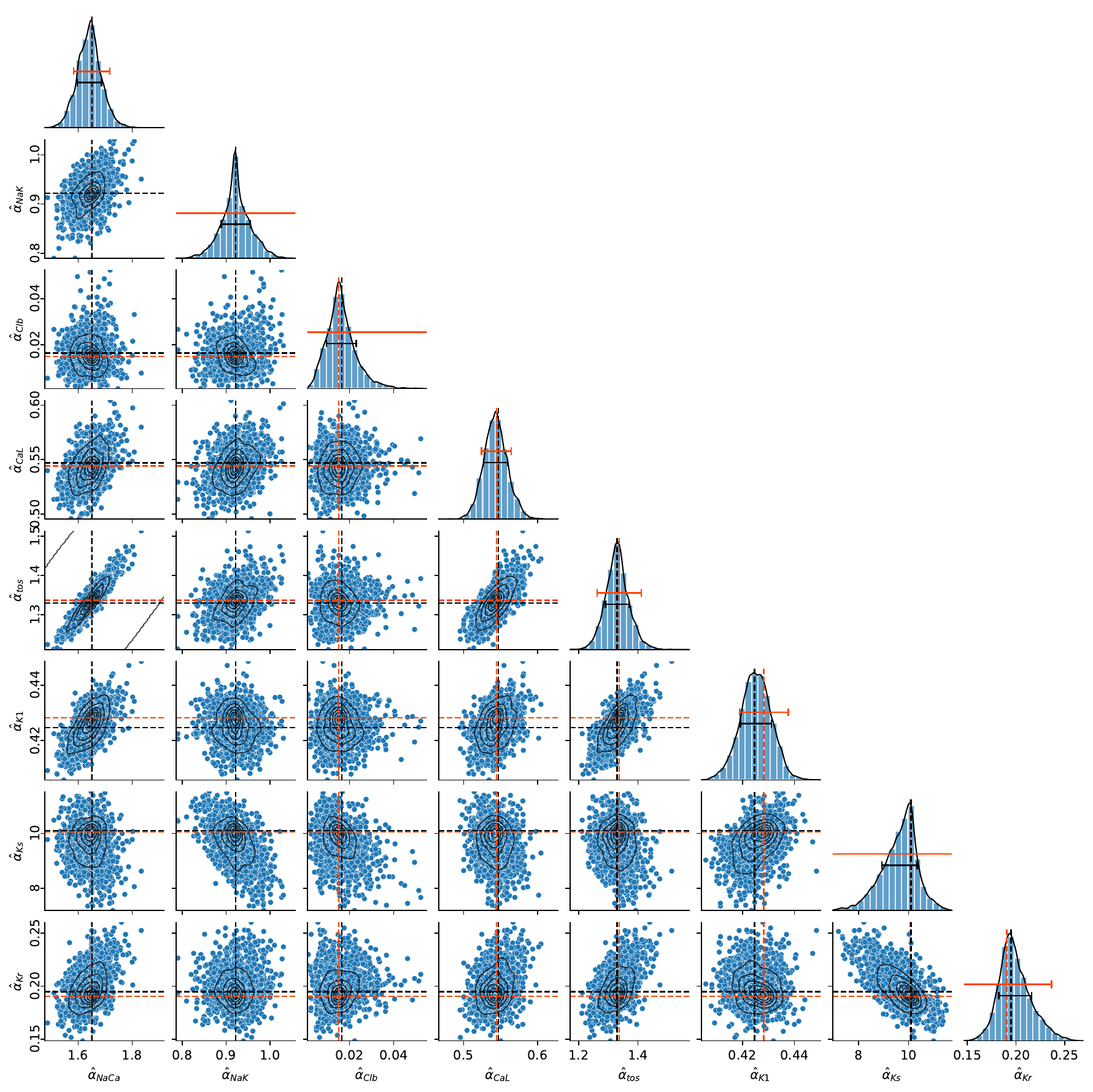}
  \put(1,93)    {\rotatebox[origin=c]{90}{{\small density}}}
\end{overpic}
\caption{\rev{Uni-variate (on the main diagonal) and bi-variate
pairwise (lower diagonal part) marginal posterior distributions in
the estimand space of cell {\sffamily uid: 21051\_run1cell16}.
Black and orange-red dashed lines show the Bayesian sampling maximum
a-posteriori probability (MAP) estimates $\hat\alpha^\text{bs}$, and
the global likelihood maximisation estimates $\hat\alpha^\text{lm}$,
respectively. In the diagonal plots, the black error bars show the
standard deviation from mean of the posterior distributions, and
the orange-red error bars show the standard errors determined
by \eqref{eq:stderr}. Kernel density estimation contours are shown throughout.} 
\label{revfig010}
}
\end{figure*}

\subsection{\rev{Bayesian inference test}}
\label{sec:bayesian}
\rev{To further test the "frequentist" methodology of
section \ref{maxlikelihood}, we undertook Bayesian inference for the
cells included in Figure \ref{fig010}. The Bayesian method provides more
accurate values for the standard errors of estimation and insights
into parameter uniqueness and identifiability, as well. 
}
\revii{An example of a similar application of both maximum-likelihood and
Bayesian methodologies to canine action potential models is presented
in \cite{Johnstone2016}.}

\paragraph{\rev{Bayesian inference.}}
\rev{The Bayesian approach assumes that, instead of being deterministic, the estimands are random variables with probability distributions given by Bayes' Theorem
\begin{gather}
P(\theta | Y) = {P(Y |\theta) P(\theta)}/{P(Y)}.
\label{eq:bayes}
\end{gather}
Here a ``prior'' probability distribution on the parameters
$P(\theta)$ is assumed from pre-experiment considerations, then
modified by the likelihood function $P(Y|\theta)$ of equation 
\eqref{eq:likelihood} and normalised by the available experimental ``evidence''
distribution $P(Y)$ to obtain the ``posterior'' probability distribution $P(\theta| Y)$. 
The latter fully describes the estimands and can be used to evaluate
any requited moments. In particular, the mode and the standard
deviation of the posterior can be compared 
to the frequentist ``best'' estimate \eqref{eq:argmax} and standard
error of estimation \eqref{eq:stderr}, respectively. 
}

\paragraph{\rev{Sampling.}} \rev{The posterior is rarely available in
closed form, but it can be sampled numerically. We used the
Haario-Bardenet adaptive covariance Markov-chain Monte-Carlo
method \citep{Bardenet2015} as implemented in \citep{Clerx2019}; this
is an algorithm for sampling involved, high-dimensional
distributions. For each cell 5 Markov chains were initialised at the
best values found by optimisation with addition of a small amount of
independent zero-mean Gaussian noise. Chains were run for 10000 steps
converging to an average scale reduction factor $\hat R$ of 1.08 and
producing 50,000 samples for each parameter as illustrated in
Supplementary Figure \ref{revsupfig02}. 
}

\paragraph{\rev{Bayesian sampling test results.}}
\rev{
Figure \ref{revfig010} illustrates the posterior probability
distribution of the estimated parameters for cell {\sffamily uid:
210511\_run1cell16} in one and two-dimensions.   
The maximum a-posteriori probability (MAP) estimates
$\hat\alpha_k^\text{bs}$ of Bayesian sampling, and
the best estimates $\hat\alpha_k^\text{lm}$ of global likelihood
maximisation agree closely for
all estimands, with the largest relative difference being
${\tilde \Delta} \hat{\alpha}_\text{Clb} =
    (\hat{\alpha}_\text{Clb}^\text{bs}-\hat{\alpha}_\text{Clb}^\text{lm})/\hat{\alpha}_\text{Clb}^\text{lm}
    = 3.2\times10^{-2}$. 
The standard deviations from the means of the posterior distributions
are smaller, often significantly smaller, than the standard errors of
estimation determined from equation \eqref{eq:stderr} as illustrated
by their comparison in the diagonal plots in
Figure \ref{revfig010}. Thus, the actual uncertainty in parameter
estimation is smaller than the rather conservative errors of
estimation $\hat\sigma_k$ that we report throughout. This is due to
the classical formula \eqref{eq:stderr} being an asymptotic
approximation strictly valid only in the limit of infinitely large
datasets $\mathcal{V}_j$. The overestimation of standard errors was
also noted in the synthetic data test of section \ref{sec:synthetic}.  
The pairwise marginal distributions of the majority of estimands show
insignificant correlations which is a strong indication that the
estimates are indeed unique. The estimands ${\alpha_\text{tos}}$ and
${\alpha_\text{NaCa}}$  are somewhat correlated with their marginal
distribution taking the form of a ridge and parameter values along
the ridge being nearly equally likely. However, we note that the
Bayesian sampling $\hat\alpha_k^\text{bs}$, and the likelihood
maximisation $\hat\alpha_k^\text{lm}$ estimates continue to agree well
in this case, too. 
The Bayesian sampling results for the rest of the cells from Figure \ref{fig010} are very similar as shown in Supplementary Table \ref{tab:reldif:BStoLM}.
}

\rev{
In summary, Bayesian sampling confirms uniqueness of estimation,
indicates a smaller parameter uncertainty than that measured by
equation \eqref{eq:stderr} and shows excellent agreement with
likelihood maximisation estimates. Because of this, and the fact that
sampling is significantly more expensive computationally, we
did not perform Bayesian analysis for the rest of the myocytes.} 

\begin{table}[t]
\begin{center}
\resizebox{\columnwidth}{!}{%
\begin{tabular}{lllllllllll}
\toprule
{} &     count &      mean &       std &       min &       25\% & 50\% &       75\% &       max &  \reviii{skewness} &  \reviii{kurtosis} \\
\midrule
$\hat{\alpha}_\text{NaCa}$ &  1180 &  7.04e$-$1 &  7.41e$-$1 &  1.09e$-$3 &  2.18e$-$1 &  4.60e$-$1 &  9.16e$-$1 &  4.79e$+$0 &  \reviii{2.07e$+$0} &  \reviii{5.11e$+$0}\\
$\hat{\alpha}_\text{NaK }$ &  1180 &  5.36e$-$1 &  6.48e$-$1 &  1.67e$-$3 &  2.27e$-$2 &  2.97e$-$1 &  8.44e$-$1 &  5.22e$+$0 &  \reviii{1.98e$+$0} &  \reviii{6.72e$+$0}\\
$\hat{\alpha}_\text{Clb }$ &  1180 &  2.24e$-$1 &  3.02e$-$1 &  6.22e$-$4 &  1.97e$-$2 &  8.45e$-$2 &  3.42e$-$1 &  2.83e$+$0 &  \reviii{2.42e$+$0} &  \reviii{9.90e$+$0}\\
$\hat{\alpha}_\text{CaL }$ &  1180 &  4.98e$-$1 &  1.40e$+$0 &  1.11e$-$4 &  1.42e$-$1 &  2.22e$-$1 &  3.68e$-$1 &  2.21e$+$1 &  \reviii{9.15e$+$0} &  \reviii{1.04e$+$2}\\
$\hat{\alpha}_\text{tos }$ &  1180 &  5.53e$-$1 &  1.43e$+$0 &  2.51e$-$4 &  7.80e$-$3 &  6.97e$-$2 &  5.41e$-$1 &  2.30e$+$1 &  \reviii{7.91e$+$0} &  \reviii{8.90e$+$1}\\
$\hat{\alpha}_\text{K1  }$ &  1180 &  4.17e$-$1 &  1.47e$-$1 &  5.33e$-$2 &  3.19e$-$1 &  3.82e$-$1 &  4.86e$-$1 &  1.01e$+$0 &  \reviii{1.01e$+$0} &  \reviii{1.11e$+$0}\\
$\hat{\alpha}_\text{Ks  }$ &  1180 &  8.45e$+$0 &  1.53e$+$1 &  1.94e$-$2 &  2.79e$-$1 &  1.08e$+$0 &  9.84e$+$0 &  1.00e$+$2 &  \reviii{2.82e$+$0} &  \reviii{9.28e$+$0}\\
$\hat{\alpha}_\text{Kr  }$ &  1180 &  3.68e$-$1 &  3.91e$-$1 &  1.07e$-$4 &  2.87e$-$2 &  2.68e$-$1 &  5.73e$-$1 &  2.54e$+$0 &  \reviii{1.56e$+$0} &  \reviii{3.46e$+$0}\\
\bottomrule
\end{tabular}
}
\end{center}
\caption{\label{tab010} Sample statistics of the population of
  parameter estimates $\mathcal{M}$. Percentages in the column
  headings denote quartiles of the data. Scientific ``e''--notation is
  used for the values of real numbers.}
\end{table}

\subsection{Cell-specific model population as a random
  sample of a ``healthy myocyte'' phenotype}
\label{sec:cellpopulation}
The action potential waveforms recorded from 1228 rabbit
ventricular myocytes were fitted. Out of these, 1180 models with goodness-of-fit
$p > 0.3$ were accepted. The value $\gamma=0.3$ is selected by
comparison with the goodness-of-fit values of the fits shown in
Figure \ref{fig010} which we consider to be good. 
Supplementary Figure \ref{supfig010} illustrates the fits
for all accepted fits in a format identical to that of
Figure \ref{fig010}.
\rev{The rejected fits were approximately 4\% of the population size. The
experimental AP traces of the rejected fits featured more pronounced
versions of the waveforms of cell {\sffamily uid: 220824\_run1cell6}
and {\sffamily uid: 220824\_run1cell20} from Supplementary Figure \ref{supfig010}
and were not captured well by the fitting procedure.}   
We now characterise the population $\mathcal{M}$ of parameter
estimates as a whole. The elements $\hat \alpha_k$ of
$\mathcal{M}$ are eight dimensional and consequently visualisation
and interpretation is challenging. 

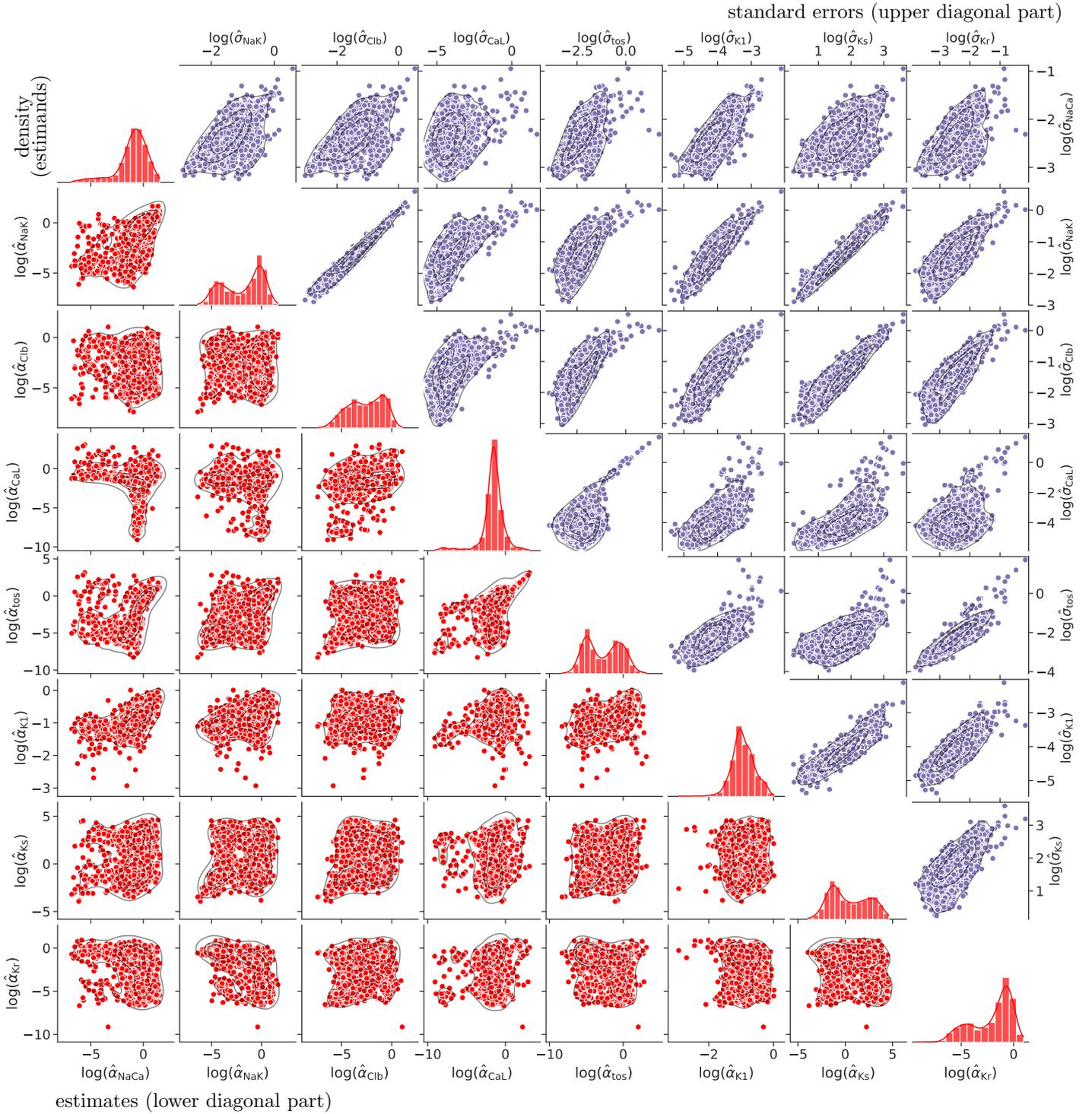
\begin{figure*}[t]
\hspace*{-10mm}\begin{overpic}[width=1.10\textwidth]{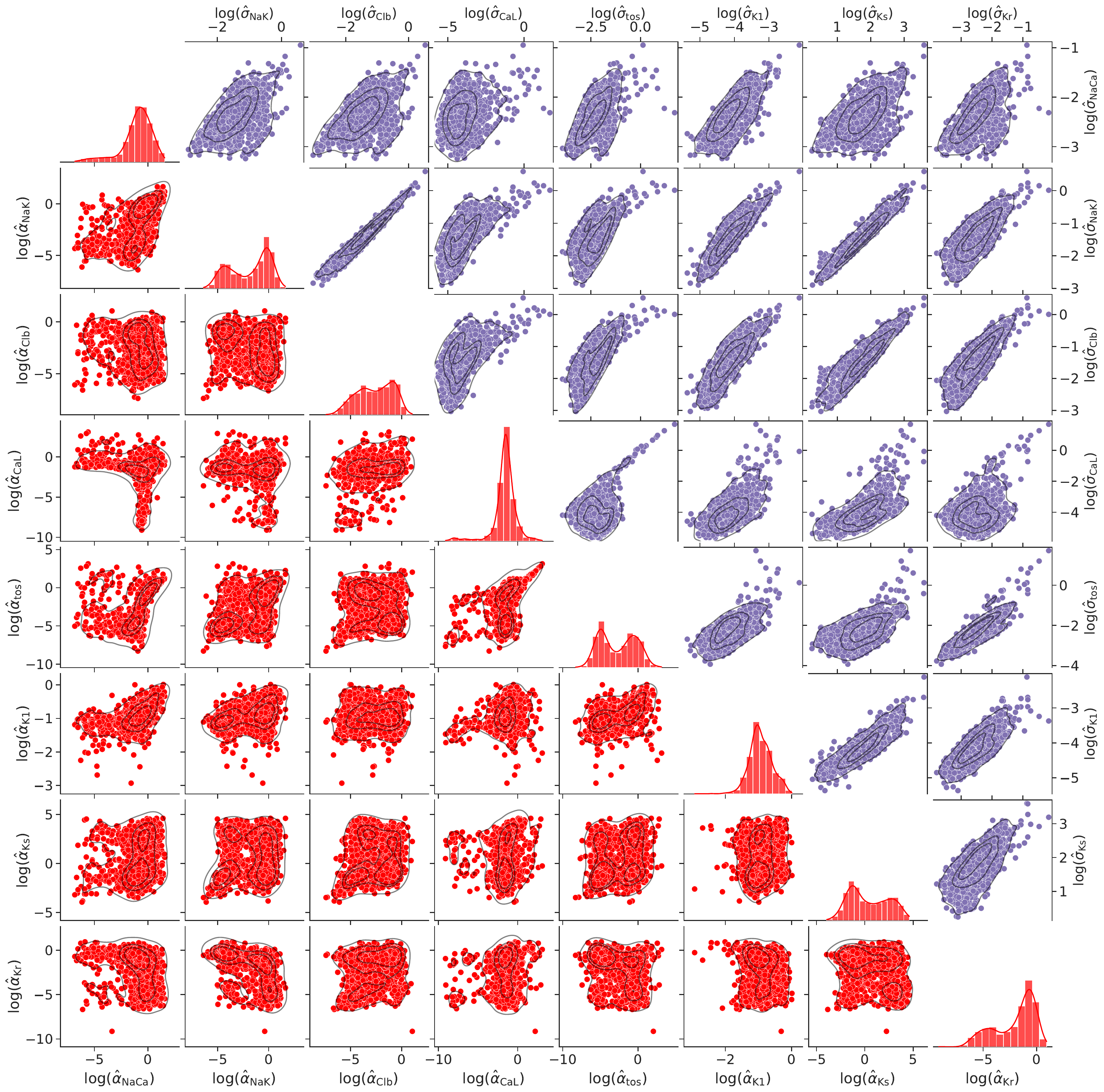} 
  \put(1.5,89)  {\rotatebox[origin=c]{90}{{\small\rev{density}}}}
  \put(3,89)  {\rotatebox[origin=c]{90}{{\small\rev{(estimands)}}}}
  \put(5.3,-2)  {\small\rev{estimates (lower diagonal part)}}
  \put(67.5,99.6)  {\small\rev{standard errors (upper diagonal part)}}
\end{overpic}
\medskip
  \caption{Marginal probability distributions of \rev{parameter estimands
    $\hat{\alpha}_k$ and their estimation errors $\hat{\sigma}_k$}. Pairwise scatter-plots     of the
 \rev{estimates $\hat{\alpha}_k$ and of their
    standard errors $\hat{\sigma}_k$ are plotted below and above the grid diagonal in red and in magenta, respectively,}
  for all 1180 accepted myocyte fits.
    Single-parameter histograms are shown on the
    grid diagonal in red. Associated marginal kernel density estimations
    are included throughout. Logarithmic scales are used.
  \label{fig040}  }
\end{figure*}

\paragraph{Population of models as a random sample of a random variable.}
We consider myocytes $C_i$ as elements of a sample space $\Sigma$ of
``normal healthy'' cells, and we consider their corresponding Shannon model
parameters $\hat \theta$ as elements of a measurable space
$\Theta$. Then, the set of parameter estimates $\mathcal{M}$ represents
a random sample from the probability distribution
$P(\theta)$ of the random variable $\Sigma:\mathcal{F} \mapsto
\Theta$, while the parameter estimation process plays the role of the
map $\mathcal{F}$.

\begin{figure*}[t]
  \begin{center}
\begin{overpic}[width=0.49\textwidth]{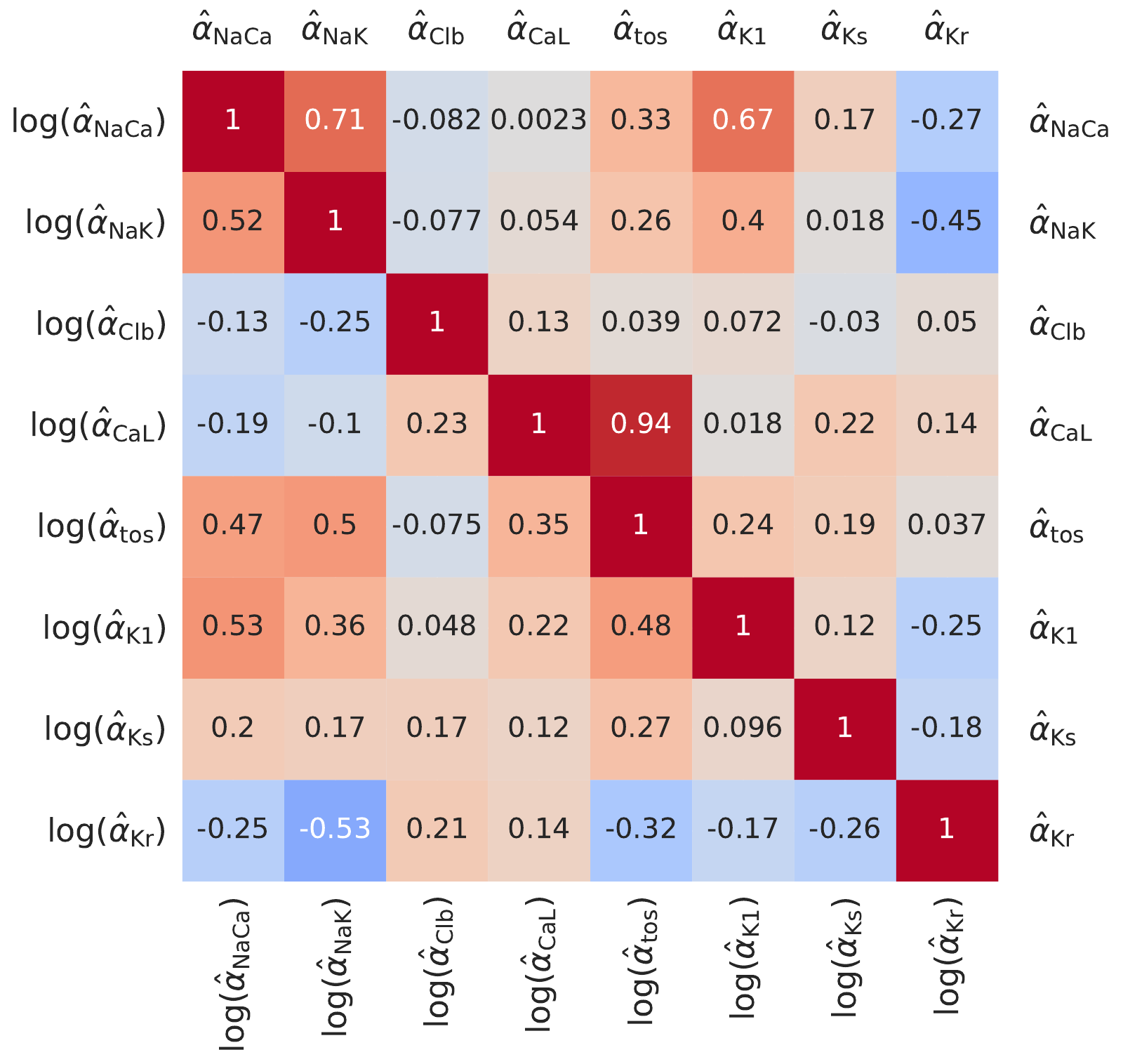}
  \put(2,97) {{\small(a) \rev{Correlations between estimands}}}
\end{overpic}
\begin{overpic}[width=0.49\textwidth]{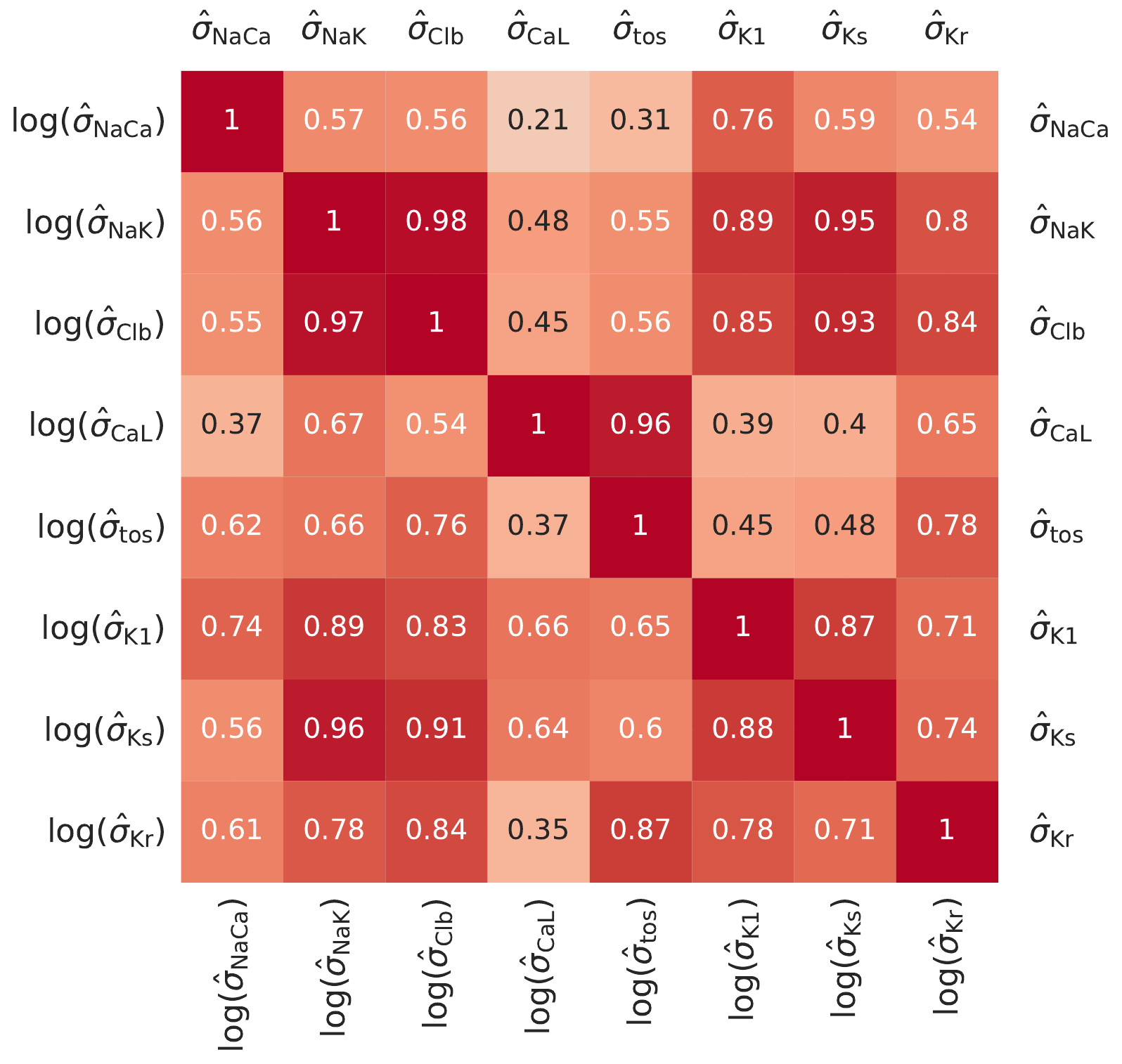}
  \put(2,97) {{\small(b) \rev{Correlations between standard errors of estimation}}}
\end{overpic}
\end{center}
  \caption{\label{fig050}
    Pairwise sample correlation coefficients $r$ for (a)
    \rev{the  parameter estimands $\hat{\alpha}_k$}, and (b) \rev{the standard
    errors in their  estimation $\hat{\sigma}_k$}.
    For convenience, \rev{coefficients of logarithmically transformed and non-transformed
    samples are plotted in the lower triangular and the upper
    triangular part, respectively}, in both panels.} 
\end{figure*}

\paragraph{The cell-specific model population.}
Attempts to characterise the probability distribution $P(\theta)$ of
this random variable with the help of the sample $\mathcal{M}$ are 
presented in Table \ref{tab010} and Figures \ref{fig040}
and \ref{fig050} in ``zero'', one and two dimensions, respectively.
Basic summary statistics including ranges, mean, standard deviations, and quartiles
are listed in  Table \ref{tab010} for each of the eight parameter estimands.
We find that estimates have large ranges of variation with
$\hat\alpha_\text{Ks}$ exhibiting a range of four orders of
magnitude. This is in stark contrast to the narrow ranges of 
variation from the baseline values typically assumed in the
literature, e.g.{} see \citep{Lachaud2022}.
We use logarithmic scales in the subsequent figures to capture these
wide ranges visually and provide a balanced view of the data.

The central tendency, dispersion and range of values provided in Table
\ref{tab010} are far from sufficient to capture the complexity of the
dataset $\mathcal{M}$. Uni-variate marginal distributions of all Shannon
parameter estimands are shown in the diagonal panels of Figure
\ref{fig040} in the form of histograms and gaussian kernel density
estimations. With the exception of $\hat\alpha_\text{K1}$,
$\hat\alpha_\text{CaL}$ and $\hat\alpha_\text{NaCa}$, which have long tails
of outliers towards small values, the distributions of all estimands
exhibit pronounced bimodality. \reviii{Skewness and kurtosis values
are also provided in Table \ref{tab010}.}

To reveal inter-variable correlations and clustering, bi-variate joint
marginal distributions are visualised in Figure \ref{fig040} in the form of
scatter-plot histograms and with contours of associated
two-dimensional gaussian kernel density estimations.
These are arranged in a grid of panels where estimates for each Shannon
parameter estimand are plotted on the $y$-axes across one of the rows
of the grid as well as on the $x$-axes across one of its columns. 
This grid arrangement is commonly known as a pairplot or correlogram
and represents a comprehensive two-dimensional view of the
dataset $\mathcal{M}$. The uni- and
bi-variate joint distributions shown in Figure \ref{fig040} are
marginal as all other estimands vary simultaneously with the one, or
the pair, that is being plotted. There are only weak, if any, linear
correlations between the parameter estimands. This is further
quantified in Figure \ref{fig050}(a) by a map of pairwise 
correlation coefficients. We recall that the sample correlation
coefficient $r_{xy}$ of two random samples $x=\{x_i\}$  and
$y=\{y_j\}$ is conventionally defined as $r_{xy} = v_{xy}/(s_x  s_y)$,
where $v_{xy}$ is their sample covariance and $s_x$ and  $s_y$ are
their sample standard deviations, and quantifies the
strength and the slope of linear relationships between variable pairs.
It is not physically possible to visualise the tri-variate and
multivariate joint distributions of the estimands.

The pairplot of Figure \ref{fig040} is symmetric with respect to the
grid diagonal. We have therefore, taken the opportunity to visualise
in the same format the bi-variate joint distributions of the standard
errors of parameter inference $\{\hat \sigma_k^j\}$ corresponding
to the estimate of the parameter values $k$ for each cell $j$. These
are plotted in magenta above the main diagonal in
Figure \ref{fig040}. Uni-variate distributions of these quantities are
not presented. In contrast to the estimands, the standard errors show
significant positive linear correlations, as also quantified in Figure \ref{fig050}(b).

\begin{figure*}[t]
  \hspace*{-3mm}\includegraphics[width=1.02\textwidth]{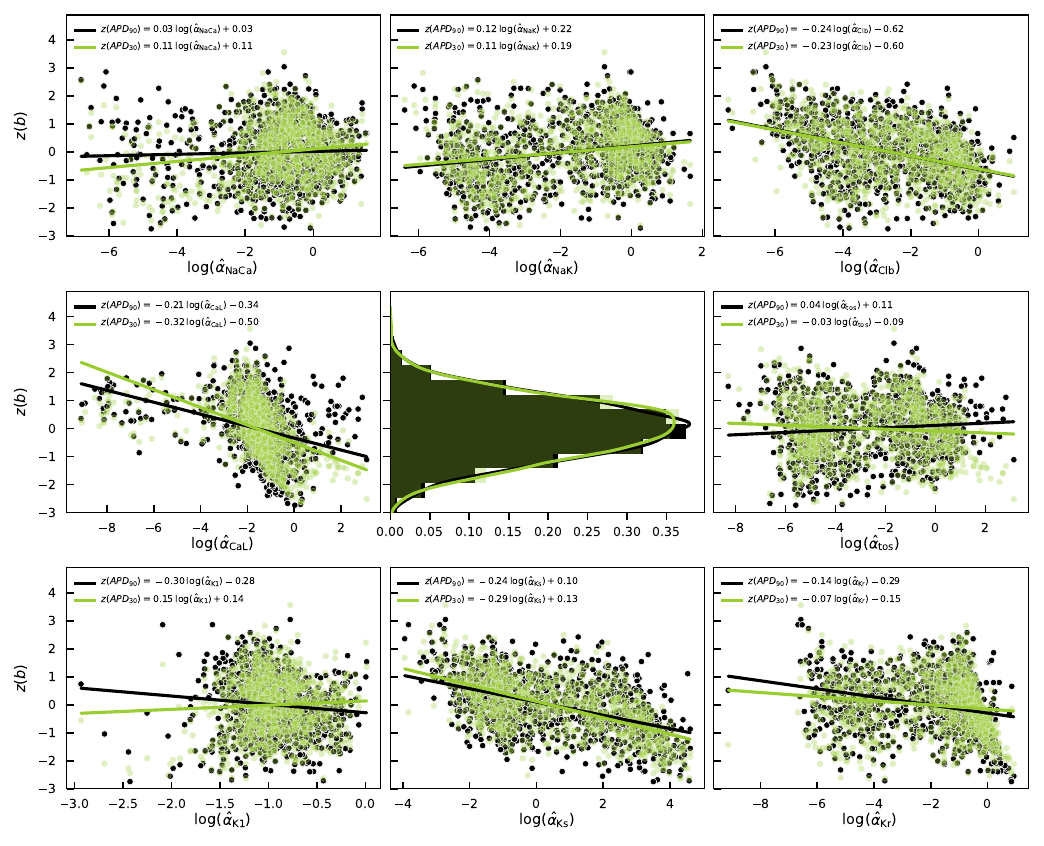}
 \caption{Action potential durations APD$_{90}$ and APD$_{30}$ as
   ``marginal'' functions of each Shannon parameter estimand
   as  indicated on the abscissas are shown in the peripheral panels
 as scatter-plots. Simultaneously, all other estimands vary randomly.
\rev{Uni-variate} linear regression fits for both biomarkers are included throughout
 and their regression coefficients are stated in the legends of each
 panel. 
Normalised histograms and kernel
   density estimates for the biomarkers are plotted in the  central panel. Histograms of the
   estimands are included in Figure \ref{fig040}.
   In all plots, the $APD_{90}$ biomarker is coloured in black, and the
   $APD_{30}$ biomarker is shown in yellow-green and made
   transparent for clarity of visualisation.    
   The biomarkers are $z$-standardised and the estimands are logarithmically transformed. 
  \label{fig060} } 
\end{figure*}

\begin{figure*}[t]
\hspace*{0mm}\includegraphics[width=1.01\textwidth,trim={0.7cm 0.8cm 1cm 0.8cm}, clip]{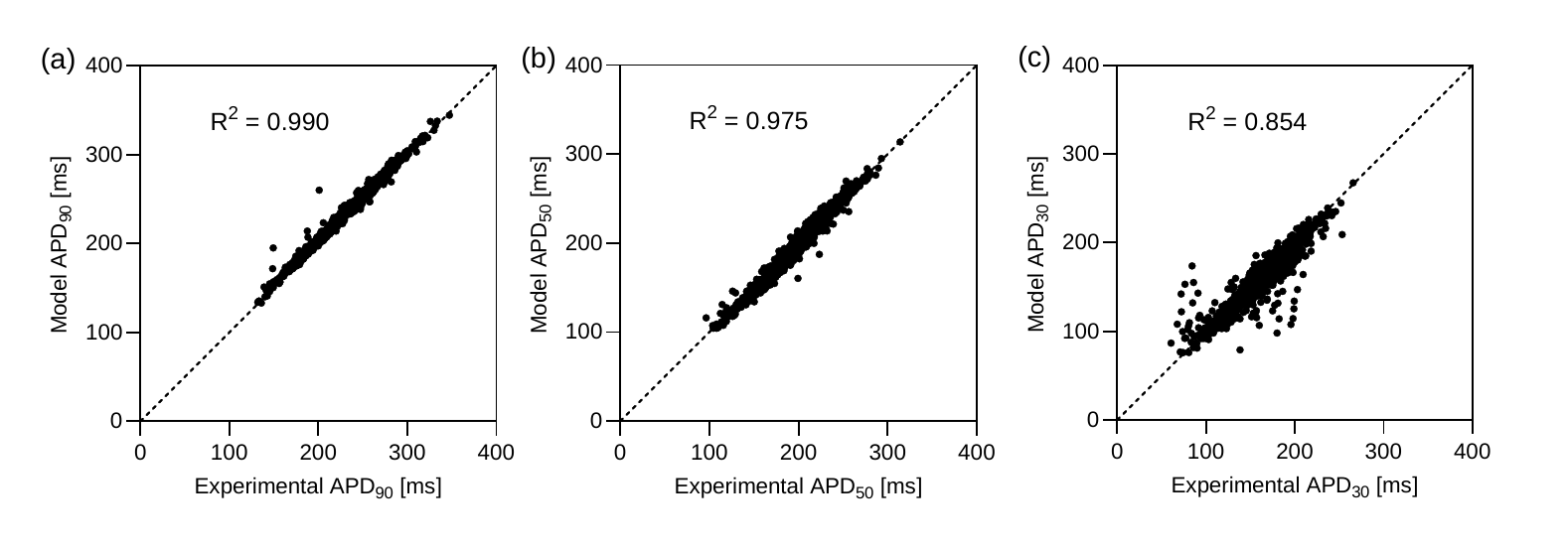}
\caption{\rev{Cell-by-cell comparison of selected biomarkers (as specified in axis
labels) computed from cell-specific Shannon models (on the ordinates)
with their experimental values (on the abscissas).
The $y=x$ dashed line denotes the position of perfect agreement.
\reviii{Values for the coefficient of determination $R^2$ for the linear regression
between the experimental values and model predictions are stated in
the legends. The p-values from a t-test for statistical significance
of the linear relationships are less than 0.001 in all cases.}}
  \label{revAPD30_APD50_APD90_ExptVsModel} } 
\end{figure*}

\paragraph{Action potential biomarkers as ``marginal'' functions of parameter estimands.}
Mathematical models relate internal parameters to physiological
observables, see discussion of Figure \ref{fig020}.
Figure \ref{fig060} provides an example of this,
now on a population level. Out of the many quantities that may
be evaluated  using the cell-specific Shannon models in the
population, we have chosen to visualise the dependence of action
potential durations APD$_{90}$ and APD$_{30}$ on each of the parameter estimands. APD$_{90}$ and
APD$_{30}$ are cellular biomarkers that are commonly
measured and reported in the electrophysiology literature.
To ensure the values of the two biomarkers are comparable, they have been
normalised by their standard $z$-score function 
$
z(b)= (b - \mu)/\sigma,
$
where $\mu$ is the mean of a sample population of random values $b$ and
$\sigma$ is its standard deviation. While not unexpected, it is
remarkable, that when standardised in this way the distributions of
both biomarkers become nearly identical.  
Similarly to the difficulties in visualising the eight-variate
probability distribution of Shannon parameter estimands, we are
restricted to presenting the parameter
dependencies of APD$_{90}$ and APD$_{30}$ in one or two
dimensions. Thus, to borrow a statistical term, the dependencies shown in Figure \ref{fig060} must be
interpreted as ``marginal'' functions in the sense that, in addition to
the parameter dependence that is explicitly plotted, all other
parameter estimands also vary simultaneously. 
Figure \ref{fig060} reveals that there are no simple functional, and
even less so, linear relationships between the two biomarkers and the
underlying parameter values. Nevertheless, we have included 
\rev{uni-variate} linear regressions fits to the panels in  Figure \ref{fig060}.
\rev{Prompted by the lack of simple \rev{uni-variate} linear relations, as
well as by prior studies in different
models \citep{Sobie2009,Sarkar2010,Morotti2017}, we have also 
computed multivariate linear regression models of the biomarkers as
a function of all eight estimands. The results along with coefficients of
determination and a comparison with the uni-variate regressions are
shown in Supplementary
Figure \ref{revsupfig_Multivariate_regression_APD30_90} and
Supplementary Table \ref{tab:multi}.
The multivariate linear regressions have coefficients of determination
$R^2 \approx 0.6$ and provide  better fits than the uni-variate linear
regressions ($R^2 \approx 0.1$). However, this is only a moderately good fit
at best and cannot serve as a replacement for cell-specific Shannon models
which provide a near exact match to the experimental biomarker
values (e.g. $R^2 = 0.99$ of APD$_{90}$) as shown in
Figure \ref{revAPD30_APD50_APD90_ExptVsModel}. \reviii{P-values for all fits are listed in the figures and indicate high statistical significance.}
The cell-by-cell agreement between models and myocytes demonstrated in
Figure \ref{revAPD30_APD50_APD90_ExptVsModel} is the key advantage of
the constructed cell-specific population compared to earlier attempts
to calibrate populations of models such as \citep{Britton2013,Muszkiewicz2016,Morotti2017,Lawson2018}.
}

The ``marginal'' functional dependencies of APD$_{90}$ 
on all pairs of estimands are included in Supplementary Figure
\ref{supfig020} in the form of a grid similar to the pairplot of Figure
\ref{fig040}. Additional biomarkers, including APD$_{50}$, peak and
resting voltage values and action potential bulk integral have also
been computed for all cells and are provided in Supplementary
Table \ref{tab:biomarkers}. As expected, there is an excellent
agreement between biomarker values determined from the cell-specific
models and values measured experimentally for each individual cell as
well as on a population level as seen in Supplementary Figure
\ref{supfig030}.

\paragraph{\revii{Prediction of intracellular calcium biomarkers.}}
\revii{The contraction of a cardiac myocyte is triggered by an intracellular
rise in calcium concentration.
Intracellular calcium concentration was
not measured in our experiments.
We have predicted the values
of four intracellular calcium biomarkers. Supplementary
Table \ref{tab:biomarkers} includes the predicted values for all 1180 cells in the
population, while Table \ref{reviitab:Cabiomarkers} provides sample
statistics and ranges for the population.
The predicted values  of the end
diastolic concentration and the peak systolic concentration of [Ca$^{2+}]_\text{i}$ fall
within the experimental ranges
reported in \citep{McIntosh2000,Baartscheer2003,Brown2008}. All
predicted biomarkers are also comparable in value to those of the Shannon baseline
model, also listed in Table \ref{reviitab:Cabiomarkers}. The Shannon
baseline model is, of course, a synthesis of data and behaviour from
a number of experimental studies as detailed within \citep{Shannon2004}.
} 
Other quantities may be evaluated on demand.

\begin{table}[t]
  \begin{center}
      \resizebox{\columnwidth}{!}{%
	\revii{\begin{tabular}{lllllllllll}
\toprule
{biomarker} &     count &      mean &       std &       min & 25\% &
50\% &       75\% &       max & baseline  &   exp.~range \& [references] \\
\midrule
ED [mM] &  1180 &  1.14e$-$4 &  6.00e$-$6 &  1.00e$-$4 &  1.12e$-$4 &
1.15e$-$4 &  1.17e$-$4 &  1.52e$-$4 & 1.02e$-$4 & (8e$-$5,2.5e$-$4), \citep{McIntosh2000,Baartscheer2003}\\
PSV [mM]  &    1180 &  5.56e$-$4 &  1.90e$-$5 &  4.52e$-$4 &
5.50e$-$4 &  5.62e$-$4 &  5.69e$-$4 &  5.75e$-$4 & 5.24e$-$4 &
(4.4e$-$4, 1.6e$-$3), \citep{McIntosh2000,Baartscheer2003,Brown2008}\\
tpeak [ms] &  1180 &  3.06e$+$1 &  1.29e$+$0 &  2.81e$+$1 & 2.98e$+$1
&  3.04e$+$1 &  3.10e$+$1 &  3.76e$+$1 & 3.41e$+$1 & \\
D50 [ms]&  1180 &  1.15e$+$2 &  1.46e$+$0 &  1.03e$+$2 & 1.15e$+$2 &
1.15e$+$2 &  1.16e$+$2 &  1.19e$+$2 & 1.15e$+$2 & \\
\bottomrule
\end{tabular}}}
\end{center}
\caption{\label{reviitab:Cabiomarkers} \revii{Sample statistics for
    selected [Ca$^{2+}]_\text{i}$ biomarkers. The corresponding values for the
    baseline Shannon model and  ranges derived from experimental data with associated
    rererences are listed in the last two columns for comparison.
The following
    abbreviations used: ED -- [Ca$^{2+}]_\text{i}$ at the end of diastole, PSV
    -- peak systolic value of [Ca$^{2+}]_\text{i}$, tpeak -- time
    from stimulus to peak [Ca]$^{2+}_i$, D50 -- period of time when
    [Ca$^{2+}]_\text{i}$ remains elevated above a threshold of
    50\% recovery from the peak value to the resting value, informally
    duration at 50\% amplitude. Percentages in the column headings denote quartiles of the data.}}
\end{table}

\subsection{\revii{Uncertainty on estimates}}
\label{sec:uncertainty}
\revii{Within the limitations of our study,
each of the constructed cell-specific models closely
matches its corresponding experimental action potential waveform as extensively demonstrated in Supplementary
Figure \ref{supfig010}. However, we recognise that our estimates are
subject to further uncertainties. These include 
structural, initial condition, simulator/procedural and others uncertainties,
see \citep{Johnstone2016}. In lieu of a disclaimer, we illustrate and
attempt to quantify some of these uncertainties here.}

\paragraph{\revii{Initial conditions uncertainty.}}
\revii{A limitation of our study is that estimation is performed
on single action potential waveforms without prepacing.
Prepacing is a procedure of forcing equations \eqref{eq:odes} into a
stable limit cycle by a periodic stimulus most often in the form of
a train of short rectangular impulses. Prepacing mimics
experimental and physiological conditions and represents an attempt to avoid the
uncertainty about the initial conditions of problem \eqref{eq:odes}.
While our experimental data is indeed prepaced by 600 APs, it is
computationally prohibitive to optimise all cells with prepacing. To
assess the effect of this discrepancy, we re-fitted nine typical cells
with prepacing by 600 APs. The ratios $\lambda_i
= \hat\alpha_i^\text{prep}/\hat\alpha_i$  between estimates obtained
with prepacing $ \hat\alpha_i^\text{prep}$ and those obtained without
prepacing $\hat\alpha_i$ are listed in
Supplementary Table \ref{tab:conversionratio}. Because the nine
example cells were selected so as to have AP waveforms ranging from
relatively short to relatively long, we assume that they sample the
entire parameter space of the problem well, and we compute the
following mean ratios 
\begin{gather}
\overline{\lambda}_\text{NaCa}    =  \text{1.07e+00},~~~~~
\overline{\lambda}_\text{NaK}     =  \text{2.48e-01},~~~~~
\overline{\lambda}_\text{Clb}     =  \text{5.19e-01},~~~~~
\overline{\lambda}_\text{CaL}     =  \text{8.22e-01},~~~~~ \nonumber\\
\overline{\lambda}_\text{tos}     =  \text{3.20e+01},~~~~~
\overline{\lambda}_\text{K1}      =  \text{1.07e+00},~~~~~
\overline{\lambda}_\text{Ks}      =  \text{2.00e-01},~~~~~
\overline{\lambda}_\text{Kr}      =  \text{2.87e+00}.~~~~~
\label{eq:topre}
\end{gather}
As a first approximation, the parameter estimates without prepacing
(plotted in Figure \ref{fig040}, listed in Supplementary Table \ref{tab:estimates} and
discussed throughout the text) can be converted to ``prepaced'' estimates by
multiplying them with the ratios of equation \eqref{eq:topre}, so for
any cell $\hat\alpha_i^\text{prep}
= \overline{\lambda}_i \hat\alpha_i$ for $ i \in \{$NaCa, NaK, $\ldots,$ Kr$\}$. }

\paragraph{\revii{Procedural uncertainty.}}
\revii{ 
The mapping of fluorescence intensity to electric potential values is
justified but somewhat arbitrary. To assess the effect of this 
choice, we have refitted 125 times cell {\sffamily uid: 21051\_run1cell16} 
while mapping the values of $V_\text{rest}$ and
$V_\text{plateau}$ to random values sampled from normal distributions
with standard deviations of 1 mV from the means of $-86$ mV and 0
mV, respectively. Sample statistics for the resulting estimates are
given in Supplementary Table \ref{suprevtab:fluorecence}. The mean values of the estimates 
remain close to those found when rest and plateau voltages are mapped
to $-86$ mV and 0 mV, respectively. }

\paragraph{\revii{Structural uncertainty.}} \revii{The Shannon \citep{Shannon2004}
model is an oversimplification of a real rabbit ventricular myocyte.
Indeed, ion channel structures and kinetics are still under study. Currents can be
modelled by various alternative approximations, e.g.{} Ohmic,
Goldman-Hodgkin-Katz, Markovian. Intracellular processes may be
described by lumped-compartment or spatially-extended sub-models, etc.
To quantify such structural uncertainty one may estimate parameters of an
alternative model. In the present case, such exercise is  of
limited validity because the alternative detailed rabbit ventricular
myocyte model, that of \textcite{Weiss2008}, is a direct extension of \citep{Shannon2004}.
}

\revii{In the light of this discussion, no claim is made about
the uniqueness of our estimates.}

\section{Conclusion}
\label{sec06}

\paragraph{Summary.}
Advances in optics-based techniques for cardiac electrophysiology
have made it possible to  develop high-throughput
platforms capable of recording transmembrane voltage from several
thousand uncoupled cardiomyocytes per hour
\citep{Mllenbroich2021}. Recent experiments based on these techniques 
reveal significant heterogeneity in uncoupled healthy myocytes both
between hearts as well as from identical  regions within a
single heart \citep{Lachaud2022}. 
In contrast, the mathematical modelling of electrophysiological
variability lags behind. Models of cardiomyocyte action potentials
describe generic cell archetypes and do not capture inter-cell variability.
This makes them ill-suited for direct use as digital twins or for
safety-related applications such as pharmaceutical drug discovery and
toxicity assessment.
To address this issue, we created a population of nearly 1200
individualised cell-specific mathematical models capable of
reproducing transmembrane potentials experimentally 
measured from healthy rabbit ventricular myocytes. 
We started from the model of \textcite{Shannon2004}, a well-regarded
and detailed mathematical model of the ionic currents in a 
generic rabbit ventricular myocyte.
We selected eight of the parameters of the model as ones most likely to affect the action
potential shape following \citep{Lachaud2022}.
We estimated cell-specific values of the eight selected parameters by
fitting voltage values computed from the Shannon model to the noisy
experimental trace from each of the biological cells.
We assumed that errors in the experimental 
measurements were normally distributed about the true signal.
We formulated a corresponding likelihood function to measure the
probability of obtaining specific experimental measurements at
particular parameter values. We then invoked the maximum likelihood
principle to find point estimates of the parameters  of interest, 
measured the standard errors of estimation and quantified the overall
goodness of fit.
We used the covariance matrix adaptation
evolution strategy, a gradient-free random search algorithm
\citep{Hansen2003}, to find a global maximum of the likelihood.
We validated the methodology by refitting synthetic data precomputed
at known parameter values, and then described in detail the fitting of 
nine typical experimental measurements.
\rev{We also tested the approach by performing Bayesian inference
which allowed us to assess the uniqueness of estimates and the size of the
estimation errors.}
We proceeded to apply the approach to action potential waveforms
recorded from 1228 rabbit ventricular myocytes using a voltage-sensitive
fluorescent indicator. We accepted 1180 fits as sufficiently good and thus
obtained a large population of cell-specific Shannon models where each
model reproduces accurately the measured electrophysiological response
of an individual cell. We interpreted this population as a random
sample from the phenotype of normal healthy rabbit myocytes.
We then attempted to characterise the probability distribution of the
phenotype by calculating basic summary statistics, and visualising
all uni-variate and bi-variate marginal distributions for the
constructed sample of parameter estimates.
A population of cell-specific mathematical models may have a large
number of diverse applications. As a simple demonstration, we computed
ionic current densities for a small subset of cells, as well as a
number of biomarkers commonly measured in experiments, including
action potential durations APD$_{30}$ and APD$_{90}$ and revealed
their dependencies on the internal state of the cells as quantified by
their Shannon model parameters.

\paragraph{Headline \rev{results}.}
\rev{In comparison with earlier studies that investigate cellular
electrophysiological variability by calibrating populations of models
and applying parameter identification techniques, the 
cell-specific Shannon models  reported here not only match
experimentally measured biomarker ranges and distributions on a
population level, but also replicate experimental biomarker values on
a cell-by-model basis.} 
Our work confirms that it is possible to efficiently and accurately
estimate model parameters at scale. We find that model parameter
distributions vary over large ranges and that parameter values are
weakly inter-correlated. As a result high-level summary
observables such as action potential duration do not depend strongly
on any one particular cellular property of the myocyte, or associated
mathematical model parameter.

\paragraph{Limitations, extensions and future directions.}
The methodology and the  applications presented here can be extended
and refined in a number of directions.
\rev{A current limitation of our study is that optimisation is
performed on single action potential waveforms, as it is
computationally prohibitive to optimise for trains of
paced action potentials. The study is restricted to stimulation at 2
Hz and the action potential dynamics at other pacing rates was not studied.} 
\rev{Bi-phasic stimuli will be considered in future refinements of the
study to better approximate the field stimulation
protocol used in experiments. The parameter $E_\text{Na,SL}$ can be included in the
optimisation to better reflect the lack of pronounced spikes in the
experimental waveforms.}
\revii{Further investigation is required into the choice and number of
fitting parameters. We are presently undertaking a global sensitivity
analysis of the Shannon model to establish a formal order of its most
sensitive parameters.} Ideally, it is desirable to fit all of
the nearly two hundred parameters of this model.
While all accepted fits accurately reproduce their corresponding
experimental measurement, it is not certain that the estimated
parameter values are the only possible ones. \reviii{Generally, the
parameter estimation problem lacks a unique solution, but
closer-to-reality estimates can be achieved by incorporating
supplementary data. Examples include complex pacing protocols,
experimental APD restitution assessment, voltage-clamp measurements,
simultaneous calcium transient recordings via microfluorimetry, or
applying selective ion channel modulators such as E4031 for hERG and
benzamil for NCX. This approach is exemplified by \textcite{Zhang2024}
who demonstrated feasibility using two separate APs per cell. 
}
Alternative models of the action potential exist for most generic
cell types, including e.g.{} the model of \textcite{Weiss2008} for the
rabbit ventricular myocyte. It is straightforward to adapt our
parameter estimation procedure for alternative models and appropriate model
selection criteria should be investigated.
From a technical viewpoint, a large number of alternative optimisation
methods exist, including both gradient-descent and random search
methods.
Whether these modifications will result in increased
accuracy and efficiency, and whether this level of detail is needed on
a population level, respectively, should be avenues for further study. 
From an electrophysiology viewpoint, it will be important to
constrain and/or extend the inference procedure by additional
experimental measurements. For example, measurements of myocyte
contraction may be incorporated by coupling the action potential model
to an appropriate model of cell contractility as performed
by \textcite{Huethorst2021}.  
\rev{Machine learning methods of the type developed
in \citep{Aghasafari2021} can be combined with the parameter
probability  distributions constructed here to identify the underlying
electrophysiology of various cell sub-populations.}
Another direction, that we plan to follow
most immediately, is to  study the pharmacodynamics of anti-arrhythmic
drugs. We have performed measurements of the response of all cells
reported in this work under the action of various concentrations of
dofetilide. Paired action potential waveforms before and after drug
administration are available for each cell. These will be used to infer
dofetilide pharmacodynamics assuming the internal state of the myocyte
can be accurately determined by the methodology developed here, or
alternatively to further constrain the parameter estimation procedure
assuming dofetilide pharmacodynamics is well-known.
\rev{This and equivalent data for a different ion channel blocker will
be part of a comprehensive examination of the action of drugs that
affect repolarisation and the subject of a future publication.} 
Possibilities for further applications are numerous and we invite the readers to
make use of the open-source software \citep{Simitev2024Code} provided with this work and
conduct their own investigations.

\begin{plain}
\section*{Statements}
\newcommand{\ethics}[1]{\paragraph*{Ethics.} #1}
\newcommand{\dataccess}[1]{\paragraph*{Data and software availability.} #1}
\newcommand{\competing}[1]{\paragraph*{Conflict of interest.} #1}
\newcommand{\funding}[1]{\paragraph*{Funding.} #1}
\end{plain}

\begin{journal}
\enlargethispage{20pt}
\end{journal}

\ethics{All procedures involving animals were performed under project licence (PP5254544) and in accordance with the UK Animals (Scientific Procedures) Act 1986.}

\dataccess{
Codes and data are available from
\href{https://doi.org/10.5281/zenodo.11191649}{doi.org/10.5281/zenodo.11191649},
see \citep{Simitev2024Code}.
}

\funding{The work of RDS and GLS was supported by the UK Engineering and Physical
Sciences Research Council [grant numbers EP/S030875/1 and EP/T017899/1].
RJG was supported by a BHF PhD studentship [grant number FS/19/56/34893].
}

\bibliographystyle{apalike3}

\end{document}